\documentclass[final,5p,times,twocolumn]{elsarticle}

\usepackage{amssymb}
\usepackage{amsmath}
\usepackage{gensymb}
\usepackage[T1]{fontenc}
\usepackage[utf8]{inputenc}

\usepackage{lineno}

\usepackage{graphicx}
\usepackage{dcolumn}
\usepackage{bm}
\usepackage{booktabs}
\usepackage{siunitx}
\usepackage{ulem}

\usepackage[colorlinks=true,linktocpage=true]{hyperref}
\usepackage{setspace}
\usepackage{comment}
\usepackage{algorithm}
\usepackage{algpseudocode}
\usepackage{bm}
\usepackage{mattens}
\usepackage{nicematrix}

\usepackage{avant} 
\usepackage{cmbright}

\newcommand{\B}{\mathbf{B}}

\newcommand {\TT}[1]{{\mathbf{#1}}}
\newcommand {\T}[1]{{\mathbf{#1}}}

\newcommand{\mub}{\boldsymbol{\mu}}

\begin{document}

\begin{frontmatter}
\begin{keyword}
    Crystal Structure \sep
    Atomic Descriptors \sep
    Supervised Learning \sep
    Molecular dynamics 
\end{keyword}

\title{Robust crystal structure identification at extreme conditions using a density-independent spectral descriptor and supervised learning}

\author[1,2]{P. Lafourcade\corref{cor1}}
\ead{paul.lafourcade@cea.fr}

\author[1,2]{J.-B. Maillet}
\author[1,2]{C. Denoual}
\author[3]{E. Duval}
\author[4]{A. Allera}
\author[4]{A. M. Goryaeva}
\author[4]{M.-C. Marinica}

\address[1]{CEA DAM DIF, 91297 Arpajon, France}
\address[2]{Université Paris-Saclay, LMCE, 91680 Bruyères-le-Châtel, France}
\address[3]{Laboratoire d'Acoustique de l'Université du Mans (LAUM), UMR 6613, CNRS, Le Mans Université, Le Mans, France}
\address[4]{Universit\'e Paris-Saclay, CEA, Service de recherche en Corrosion et Comportement des Matériaux, SRMP, F-91191 Gif-sur-Yvette, France}

\date{\today}
\begin{abstract}
The increased time- and length-scale of classical molecular dynamics simulations have led to raw data flows surpassing storage capacities, necessitating on-the-fly integration of structural analysis algorithms. 
As a result, algorithms must be computationally efficient, accurate, and stable at finite temperature to reliably extract the relevant features of the data at simulation time.
In this work, we leverage spectral descriptors to encode local atomic environments and build crystal structure classification models. In addition to the classical way spectral descriptors are computed, i.e. over a fixed radius neighborhood sphere around a central atom, we propose an extension to make them independent from the material's density.
Models are trained on defect-free crystal structures with moderate thermal noise and elastic deformation, using the linear discriminant analysis (LDA) method for dimensionality reduction and logistic regression (LR) for subsequent classification. 
The proposed classification model is intentionally designed to be simple, incorporating only a limited number of parameters. This deliberate simplicity enables the model to be trained effectively even when working with small databases. Despite the limited training data, the model still demonstrates inherent transferability, making it applicable to a broader range of scenarios and datasets.
The accuracy of our models in extreme conditions (high temperature, high density, large deformation) is compared to traditional algorithms from the literature, namely adaptive common neighbor analysis (a-CNA), polyhedral template matching (PTM) and diamond structure identification (IDS). 
Finally, we showcase two applications of our method: tracking a solid-solid BCC-to-HCP phase transformation in Zirconium at high pressure up to high temperature, and visualizing stress-induced dislocation loop expansion in single crystal FCC Aluminum containing a Frank-Read source, at high temperature. 
\end{abstract}

\end{frontmatter}

\section{Introduction}
\label{sec:introduction}
In the last decades, atomic scale simulations such as \textit{ab-initio} calculations and molecular dynamics (MD) have been increasingly used to model materials properties based on atomic-scale processes.
As large-scale simulations are needed to realistically simulate the dynamics of extended systems (e.g. linear defects, interfaces) over long simulated times, MD simulations have been successfully used to investigate a wide range of thermodynamic regimes due to their favorable linear scaling of computational time vs. system's size.
More recently, MD simulations based on empirical potentials have seen a considerable increase of their accuracy and accessible time- and length-scale, as they benefited from the development of \textit{ab-initio} calculations and machine learning (ML) methods --resulting in more accurate force fields, in conjunction with the development of high performance computing (HPC) facilities and more efficient computational methods.
As a result, simulations based on \textit{ab-initio}-accurate ML force fields have reached the scale of tens~\cite{nguyen2021billion,guo2022extending} to hundreds~\cite{johansson2022micron} of billions atoms, allowing studies to be carried out at the micrometer scale where direct microstructure comparison with experiments becomes possible~\cite{Zepeda2017}.
However, trajectories of atoms obtained via atomistic simulations must be further processed in order to extract the properties and statistics of objects of interest, i.e. defects, phases, interfaces or precipitates for example.

Present-day standard visualization methods can help to identify structural changes by analyzing atomic trajectories. However, the automation of such quantitative analysis as well as a robust identification and extraction of crystalline defects is still challenging. 
For this purpose, numerous computational methods have been developed in order to enable post-processing analysis of particle-position datasets.
These methods generally proceed by comparing individual atomic environments to that of a reference structure, while allowing a certain tolerance in the result.
The most common methods for the local structure analysis include basic energy thresholding, censtrosymmetry parameter (CSP) analysis~\cite{Kelchner1998}, bond order analysis~\cite{Steinhardt1983}, common neighbor analysis (CNA)~\cite{Honeycutt1987, Faken1994}, adaptative common neighbor analysis (a-CNA)~\cite{Stukowski2012}, bond angle analysis (BAA)~\cite{Ackland2006}, Voronoi analysis \cite{Lazar2015}, neighbor distance analysis (NDA)~\cite{Stukowski2012} and polyhedral template matching (PTM) analysis~\cite{Larsen2016}. 

Another set of techniques, oriented towards continuum mechanics measures, has been proposed in the literature to identify the deformation state as well as defects such as dislocation lines. In order to observe dislocation-mediated plasticity, such tools can be used to filter the data, remove crystalline atoms, and extract the atoms that constitute the core of defects. However, no information can be extracted concerning the type of crystal defects, which can be represented by vacancies, interstitials, or dislocation lines. Also, atoms identification can fail, and cases where defects overlap (e.g. an interstitial is found in a dislocation core) is not taken into account.
Continuum-like measures, computed as per-atom variables over the current neighborhood, with respect to a reference configuration, have also been introduced \cite{Tucker2009, Tucker2011, Zimmerman2009, Hartley2005, Cermelli2001}, allowing the computation of local deformation gradient tensors, slip vectors or Nye tensor. The combination of the latter techniques, along with structural identification methods, have also been used to identify dislocation lines and their Burgers vectors. However they become inefficient at high temperatures or when dislocation interactions come into play. Another widely-used method, known as the dislocation extraction algorithm (DXA)~\cite{Stukowski2010dxa, Stukowski2012dxa} allows to consecutively mesh the atomistic configuration, map the local tetrahedra to perfect crystal structures, extract the distorted tetrahedra (i.e. disordered atoms), and perform the Burgers circuit in order to extract the dislocation lines as well as their Burgers vectors. This algorithm is very helpful for monitoring the evolution of dislocation densities over time~\cite{Bulatov2006, Zepeda2017, Zepeda2020} and characterizing crystal plasticity features~\cite{Bertin2020}.

Most of the methods listed above are directly available in the Open Visualization Tool (\texttt{OVITO})~\cite{ovito_ref}, enabling a straightforward comparison with newly developed structure identification tools. However, a common weakness of traditional analysis methods is their sensitivity to thermal noise which can be limiting when simulating crystals at finite temperature.
To remove thermal vibrations while preserving the features of the high-temperature structure, vibration-averaging can be used, as well as structure denoising based on graph neural networks~\cite{hsu2023}.

Some of the techniques described above have been used to process atomistic simulations on-the-fly, benefiting from their ease of implementation and high data throughput since they are easily integrated in MD codes such as \texttt{LAMMPS}~\cite{Plimpton1995, Thompson2022} or ExaSTAMP~\cite{Cieren2014, Cieren2015Thesis, Prat2019Thesis, Prat2020}.
However, more computationally demanding methods, such as DXA, remain challenging to use for on-the-fly, large-scale application.
Although the DXA implementation available in \texttt{OVITO} is highly optimized, the algorithm is computationally expensive and requires 1 GB of free RAM memory per million atoms.
The memory should be distributed across several nodes to scale to large systems, or a parallelism scheme should be adopted, which is not the case in the official distribution but seems to be a work in progress on their side according to very recent communications.
For example, one cubic micrometer sample of BCC Tantalum with only the positions written to a single ASCII file would contain approximately 55 billion atoms and occupy 24 GB of disk space. In order to run DXA on this sample, up to 55,000 GB of memory would be required, which is technically difficult to achieve without special hardware design. 
This shows the urgent necessity to develop \textit{in-situ} analysis tools, since filtering such simulations on-the-fly can cut the needs of storage capacity by orders of magnitude.
In addition to basic properties, other descriptors like deformation gradient tensors, velocity-gradient tensors or bond-order parameters for example would drastically increase the necessary disk space. It becomes clear that storing a few terabytes per snapshot over a few nanoseconds, even with a low output frequency, cannot be considered as viable.

More lately, additional atomic descriptors enabling local structure analysis have been introduced \cite{Musil2021, Chung2022}, such as Behler-Parrinello Chebyshev polynomial representations (CPR)~\cite{Artrith2017}, Behler-Parrinello symmetry functions (BP)~\cite{Behler2011}, smooth overlap of atomic positions (SOAP)~\cite{De2016}, atomic cluster expansion (ACE)~\cite{Zeni2021}, adaptative generalizable neighborhod informed features (AGNI)~\cite{Batra2019,Chandrasekaran2019}. In addition, machine learning aided crystal structure identifiers have been published, either based on Bayesian deep learning (ARISE)~\cite{Leitherer2021} or neural networks~\cite{DeFever2019}. 
Finally, structural defects in crystalline solids can be effectively detected as structural outliers using distortion scores of local atomic environments~\cite{Goryaeva2020, goryaeva_compact_2023}. This method uses minimum covariance determinant (MCD) in conjunction with compact atomic descriptors like bispectrum~\cite{bispectre} and allows for accurate structural analysis even in noisy structures. However, until now, it has not been coupled with an automated structure classifier.
These methods, while being highly accurate, also have a non-negligible computational cost compared to traditional methods presented above. However, the performance of BSO4 for example, has been dramatically improved since the initial version of SNAP (see \cite{nguyen2021billion}) and was ported to GPU, making it one of the most efficient force-field framework with near-\textit{ab-initio} accuracy. For practical use in on-the-fly MD analysis, the calculation of BSO4 every Nth step should thus not be a bottleneck --even when using another force-field than SNAP. The present methodology tightly integrates with the MD engine with acceptable overhead.

In this work, we address \textit{in-situ} analysis of large-scale MD trajectories and strive to minimize the amount of information to be stored. While today the research community aims at simulating larger and larger samples through MD simulations, the bottleneck is not only in performing the simulation itself, but mainly in effectively and accurately analyzing it, which represents a paradigm shift. 
However, there is always a trade-off to find between computational cost, accuracy, and robustness since a low sensitivity to atomic displacements usually comes at the price of a reduced capability of the identification method to distinguish similar structures~\cite{Stukowski2012}
Here, we propose a novel algorithm (see Algorithm \ref{alg:classifier}) for the identification and classification of crystal structures, also allowing for the accurate detection of atoms that contribute to defects. The method uses machine learning (ML) techniques and can be used for the analysis of materials under extreme conditions, including thermal noise, large deformations as well as large hydrostatic pressures.

The paper is organized as follows. The first section details the construction of the training database used for different models. The second section describes the training process for the different algorithms employed in this work. Finally, in the last section, we demonstrate the performance of our crystal structure identification algorithm and apply it to analyze large-scale MD simulations at finite temperature. We present two cases of interest: solid-solid phase transition in hexagonal close-packed Zirconium and Frank-Read source dislocation loops expansion in face-centered cubic Aluminum.

\begin{algorithm}[H]
    \small
    \caption{Crystal structure classification algorithm}
    \begin{algorithmic}
    \State{$C:$ Number of crystal structures in the database }
    \State{$M:$ Dimension of the atomic descriptor}
    \State{$\mub_\mathrm{cs}:$ Mean descriptor of sub database with structure $\mathrm{cs}$}
    \State{$\TT{\Sigma}_\mathrm{cs}:$ Covariance matrix of sub database with structure $\mathrm{cs}$}
    \State{$\epsilon_\mathrm{cs}:$ Acceptance threshold for structure $\mathrm{cs}$}
        \For{Each atom in simulation cell}
            \State{1. Compute per-atom descriptor $\B \in \mathbb{R}^{D}$}
            \State{2. Reduce dimension $\mathbf{x} = P_{\textrm{LDA}}(\B)$ $P_{\textrm{LDA}}: \mathbb{R}^{D} \to \mathbb{R}^{C-1}$} 
            \State{3. Logistic regression $\mathrm{cs} = P_{\textrm{LR}}(\mathbf{x})$ $P_{\textrm{LR}}: \mathbb{R}^{C-1} \to \mathbb{R}^{1}$}
            \State{4. Compute $d^\mathrm{cs}_\mathrm{Maha} = \sqrt{(\mathbf{x}-\mub_\mathrm{cs})^\mathrm{T} \TT{\Sigma}^{-1}_\mathrm{cs} (\mathbf{x}-\mub_\mathrm{cs})}$ }
            \If{$d^\mathrm{cs}_\mathrm{Maha} > \epsilon_\mathrm{cs}$}
                \State{Define atom as non-crystalline, i.e. $\mathrm{cs}=-1$}
            \Else
                \State{Assign crystal structure $\mathrm{cs}$ to atom}
            \EndIf
        \EndFor
    \end{algorithmic}
\label{alg:classifier}    
\end{algorithm}

\section{Database preparation}
\label{sec:database}
The training database contains four different crystal structures: body-centered cubic (BCC), face-centered cubic (FCC), hexagonal close-packed (HCP), and cubic diamond (c-DIA). The extension to other crystalline structures is straightforward.
For each structure, a model metal is considered, and modelled using the following interatomic potentials: Aluminium~\cite{mendelev_pm_2008} for BCC, Iron~\cite{mendelev_pm_2003} for FCC, Zirconium~\cite{mendelev_pml_2007} for HCP and Silicon~\cite{stillinger_prb_1985} for c-DIA.
A common aspect of supervised machine learning techniques is that their application range is given by the information contained in the learning database. Hence the elements of the training database should be carefully chosen with respect to target applications. 
Here, we pay attention to include the structures carrying information about temperature and small deformations. Then, once built, the database is mapped into a descriptor space onto which learning will be performed. These procedures are detailed below.

\subsection{Construction of the database in Cartesian space}
\subsubsection{Finite temperature molecular dynamics trajectories }
The effects of temperature must be accounted for when developing a robust crystal structure classifier suitable for materials at extreme conditions.
In order to sample finite-temperature configurations for the database, we compute MD trajectories in the  NPT ensemble for each material representing the different crystal structures.
The simulation cell dimensions correspond to the equilibrium density of each material at 300 K and ambient pressure, and contain 864 atoms. Trajectories are integrated with a timestep of 1 fs, and the coupling parameters for the thermostat and barostat are set to 0.1 and 1.0 ps, respectively.
NPT simulations are performed at ambient pressure while ramping up the temperature from 0 to 2/3 of each material's melting temperature over a 500 ps time window. Configurations are extracted every 5 ps along each trajectory, leading to an ensemble of 100 configurations per crystal structure.

\subsubsection{Deformation measure}
A macroscopic deformation gradient tensor $\TT{F}$ is applied to the entire system while remapping the $3N$ atoms coordinates $\mathbf{q} = \T{r}_1 \oplus \ldots \oplus \T{r}_N \in \mathbb{R}^{3N}$ (where $\T{r}_i \in \mathbb{R}^3$ are the cartesian coordinates of the $i^{th}$ atom) into the deformed simulation cell. For every configuration extracted from the NPT trajectories:
\begin{equation}
    \mathbf{r}_{i, \textrm{deformed}} = \TT{F} \mathbf{r}_i,
\end{equation}
where $\mathbf{r}_{i, \textrm{deformed}} \in \mathbb{R}^{3}$ stands for the cartesian coordinates of the $i^{th}$ atom subjected to a homogeneous deformation governed by $\TT{F}$. This deformation gradient tensor reads:
\begin{equation}
    \TT{F} = \begin{pmatrix}
    F_{11} & F_{12} & F_{13} \\
    0 & F_{22} & F_{23} \\
    0 & 0 & F_{33} 
    \end{pmatrix}
\end{equation}
leading to $6$ independent variables describing all homogeneous deformation possibilities in any material. In the present work, we focus on the measure of deviatoric deformation $\epsilon_\mathrm{eq} = \sqrt{\frac{3}{2} \mathrm{dev} (\TT{E}) : \mathrm{dev} (\TT{E})}$ as a threshold criterion. Here, $\mathrm{dev}\TT{(E)} = \TT{E} - \frac{1}{3}\mathrm{tr}(\TT{E}) \TT{I}$ with $\TT{E}$ the Green-Lagrange strain tensor constructed from $\TT{F}$ with  $\TT{E}=1/2(\TT{F}^T\TT{F}-\TT{I})$ and $\TT{I}$ the identity. Only small strains, i.e. within the elastic deformation regime, are considered by imposing a threshold value on $\epsilon_{eq}$, i.e. $\epsilon_{eq}<0.05$. This way, subsets of atomistic configurations subjected to large local deformation such as dislocation-mediated plasticity or amorphous shear banding should emerge as outliers during the classification process.

\subsection{Mapping the database into the descriptor space}
\label{subsec:descriptor_space}
\subsubsection{Bispectrum SO4 descriptor}
In this work, we use the bispectrum SO4 \cite{bispectre} to map the local atomic density function into invariant representations. This descriptor has several advantages compared to using the Cartesian coordinates of the surrounding atoms, e.g., it has constant dimension and is invariant to atomic permutation, rotation, and translation. It has been widely used in the context of machine learning interatomic potentials - MLIP, including spectral neighbor analysis potentials (SNAP) \cite{snap} and other similar forms \cite{snap2, goryaeva2021, Anruo2023}. Bispectrum SO4 also has the capacity to provide an accurate description of the atomic neighborhood, suitable for advanced structural analysis \cite{bispectre,Goryaeva2020}. Below, we briefly recall the key concepts and the algebraic formalism used to compute this descriptor. The fully detailed mathematical definition is given in \cite{snap}. For all monoatomic systems employed in the present study, the atomic neighbor density around atom $i$ at location $\T{r}_i$ reads:
\begin{equation}
    \rho_i(\T{r})=\delta(\T{r})+\sum_{r_{ii'}<r_{\mathrm{cut}}} f_c(r_{ii'})\delta(\T{r}-\T{r_{ii'}})
\end{equation}
where $\T{r}_{ii'} = \T{r}_{i} - \T{r}_{i'}$ is the distance between the central atom $i$ and the neighbor atom $i'$, and the cutoff function $f_c$ ensures that the contribution of neighboring atoms smoothly decreases to zero at $r_{\mathrm{cut}}$. By mapping radial neighbor coordinates $r$ to an angular component $\theta_0=\theta_0^{\mathrm{max}} r/r_{\mathrm{cut}}$, the atomic neighbor density can be expanded in the basis functions of the unit 3-sphere, the 4D hyper-spherical harmonics $U^j_{m,m'}(\theta_0,\theta,\phi)$:
\begin{equation}
    \rho(\T{r}) = \sum_{j=0,1/2,...}^{\infty} \sum_{m=-j}^{j} \sum_{m'=-j}^{j} u^{j}_{m,m'} U^{j}_{m,m'}(\theta_0,\theta,\phi),
\end{equation}
where the expansion coefficients $u^{j}_{m,m'}$ are a sum over discrete values of the corresponding basis function evaluated at each neighbor position,
\begin{equation}
    u^{j}_{m,m'} = U^{j}_{m,m'}(0) + \sum_{r_{ii'}<r_{\mathrm{cut}}} f_c(r_{ii'})U^{j}_{m,m'}(\theta_0,\theta,\phi).
\end{equation}
Finally, using the scalar triple products of these expansion coefficients, the real-value bispectrum components can be expressed as:
\begin{equation}
    B_{j_1,j_2,j} = \sum_{m,m'} u^{j*}_{m,m'}\sum_{\substack{m_1,m'_1\\m_2,m'_2}} H\substack{jmm'\\j_1m_1m'_1\\j_2m_2m'_2} u^{j_1}_{m_1,m'_1} u^{j_2}_{m_2,m'_2},
\end{equation}
with $*$ the complex conjugation operator and where the constants $H\substack{jmm'\\j_1m_1m'_1\\j_2m_2m'_2}$ are the Clebsch-Gordan coefficients for the hyper-spherical harmonics.
The final coefficient is invariant to rotation and permutation. The order of the expansion $J_{max}$ determines the accuracy of the geometrical representation of the atomic neighborhood, although bispectrum coefficients are not listed in order of importance.
However, increasing the value of $J_{max}$ leads to better accuracy but also to a higher computational cost. In the following, we choose a value of the expansion parameter $J_{max}=4$ that represents a good compromise between the accuracy of geometrical description and computational cost \cite{goryaeva2019, Goryaeva2020}. This leads to a bispectrum $\mathbf{B}$ with $55$ real components, i.e. $\in \mathbb{R}^{55}$, which corresponds to the dimensionality of feature space.

\subsubsection{Fixed cutoff or fixed number of neighbors computation}
\label{subsec:bcut_bnn}
\begin{figure*}[t!]
    \centering
    \includegraphics[width=\linewidth]{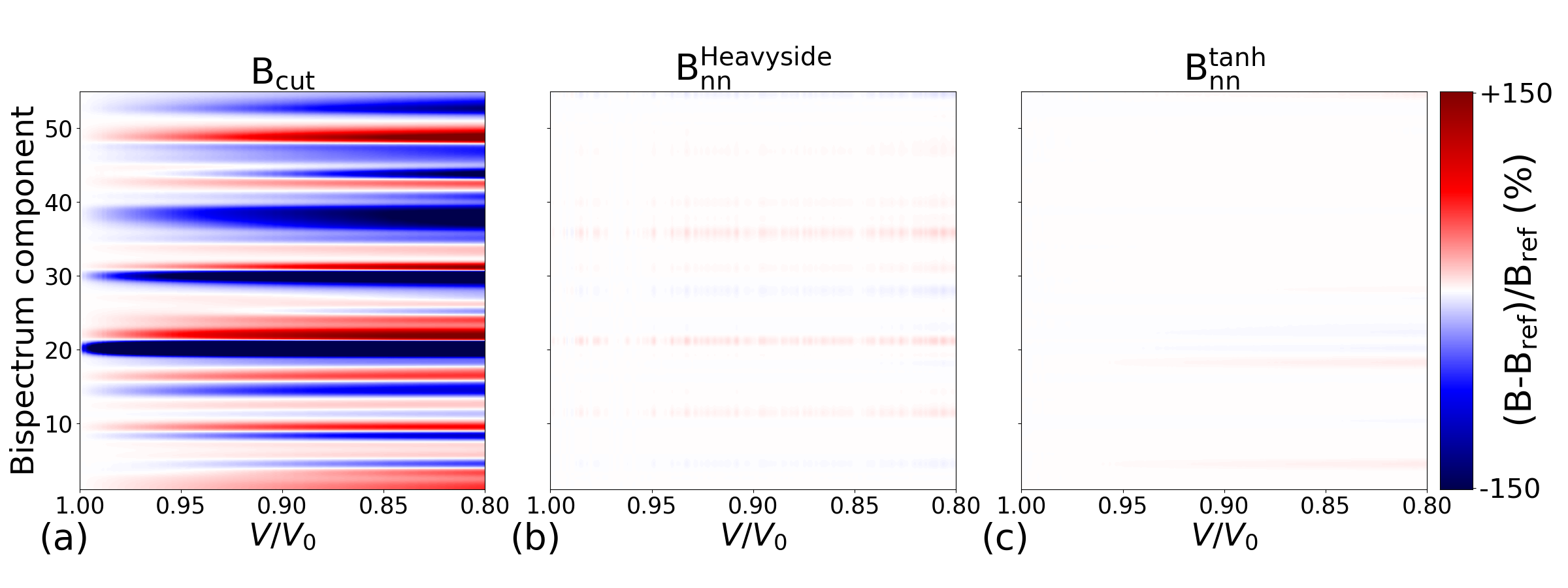}
    \caption{Construction of a density-invariant descriptor. Heatmaps showing the relative evolution of bispectrum components (in \%) as a function of volumetric ratio with respect to a reference bispectrum taken at $V/V_0 = 1.0$. Panels correspond to different bispectrum variants and setups: \textbf{(a)} Fixed \SI{6}{\angstrom} cutoff distance. \textbf{(b)} and \textbf{(c)} Fixed number of neighbors $N_\mathrm{neigh}$=24 with a Heavyside and tanh weight functions, respectively. While a fixed number of neighbors considerably reduces the density dependence, the regularized $\tanh$ weight function achieves near-invariance with respect to the local density at finite temperature.}
    \label{fig:bcut_bnn_density}
\end{figure*}

Concerning the cutoff parameter $r_{\mathrm{cut}}$ used to define the neighborhood of a central atom for which the bispectrum is computed, two strategies are proposed. Firstly we consider a fixed cutoff radius, as for the calculation of the potential which is the common procedure in the context of MLIP. %
Each neighbor in the cutoff sphere is included in the bispectrum calculation, weighted by the cut-off function which smoothly switches from 1 for distances lower than $r_{\mathrm{cut}}$ to exactly zero above $r_{\mathrm{cut}}$. 
The number of neighbors $N_\mathrm{neigh}$ of a central atom is determined by the magnitude of $r_{\mathrm{cut}}$, resulting in a larger number of neighbors for denser materials.

The second alternative is to compute the bispectrum of an atomic environment containing a fixed number of neighbors $N_\mathrm{neigh}$. Its main advantage is to remove the dependence of the crystal structure analysis on the density of the material. Hence different materials at varying densities (even locally) could be mapped to an equivalent descriptor representation. 

A simple way to achieve the selection of the $N_\mathrm{neigh}$ neighbors would be to sort them by their distance to the central atom and consider only the $N_\mathrm{neigh}$ first ones while choosing the cutoff radius as the distance between the central atom and its farthest neighbor. More formally, using a basic dichotomy algorithm, one can compute the optimal cutoff radius $r_ \mathrm{cut}$ that satisfies these requirements by defining the following equality:
 \begin{equation}
     W_i(r_ \mathrm{cut}) = \sum_j^M H(r_ \mathrm{cut} - |r_{ij}|) = N_\mathrm{neigh},
 \end{equation}
where $W_i(r_ \mathrm{cut})$ is the total weight factor associated with the central atom $i$, $H$ is the Heavyside function, $N_\mathrm{neigh}$ is the target number of neighbors and $M$ is the actual number of neighbors, related to $r_ \mathrm{cut}$, required to satisfy  $W_i(r_ \mathrm{cut})=N_\mathrm{neigh}$. Using the Heavyside weight function systematically leads to the solution with $M=N_\mathrm{neigh}$ and $r_ \mathrm{cut}$ equal to the distance to the $N^\mathrm{th}$ neighbor. However, at finite temperature, the BSO4 descriptor strongly depends on local thermal fluctuations since the position (and the identity) of the last neighbor may vary from one step to another.
A way to regularize the neighborhood construction by limiting thermal temperature effects is to use a smooth weight function such as tanh:
\begin{equation}
    W_i(r_ \mathrm{cut}) = \sum_j^M \frac{1}{2} \left(1 - \tanh{\frac{r_{ij}-r_ \mathrm{cut}}{\delta}}\right) = N_\mathrm{neigh} \; .
\end{equation}
The $\delta$ parameter ensures the smooth transition for the weights from 1 to 0 and is set to \SI{0.3}{\angstrom} in the present work, close to thermal fluctuations of atomic positions. In Algorithm \ref{alg:rho_free_bso4} we present an algorithm able to compute a general $\B$ descriptor for a constant numbers of neighbours. 
\begin{algorithm}[H]    \small
    \caption{Density-independent $\B$}
    \begin{algorithmic}
    \State{$N_\mathrm{neigh}:$ Target number of neighbors}
    \State{$W_i:$ Weight function, \textit{Heavyside} or 
    \textit{tanh}}
    \State{$r_\mathrm{neigh}:$ cut-off for initial neihgbor list}
    \State{$\mathcal{S}(i, r):$ neighbour list of the $i^{th}$ atom within distance $r$}
        \If{($W_i$ == \textit{tanh})}
            \State{Choose tolerance factor $\delta$}
        \EndIf
        \For{Each atom $i$ in simulation cell}
            \State{1. Build initial neighbor list of the i$^{th}$  atom, $\mathcal{S}(i, r_\mathrm{neigh})$}
            \State{2. Find optimal $r_\mathrm{cut}$ to minimize $\mathcal{L} = |W_i(r_\mathrm{cut})-N_\mathrm{neigh}|$}
            \For{Each neighbor $j$ in $\mathcal{S}(i, r_\mathrm{neigh})$}
                \State{Compute interatomic distance $d=\sqrt{(\mathbf{r}_j-\mathbf{r}_i)^2}$}
                \If{$(d<r_\mathrm{cut})$}
                    \State{Add neighbor $j$ to $\mathcal{S}(i, r_\mathrm{cut})$}
                \EndIf
            \EndFor
            \State{3. Basic computation of $\B$ using neighbor list $\mathcal{S}(i, r_\mathrm{cut})$}
        \EndFor
    \end{algorithmic}
\label{alg:rho_free_bso4}   
\end{algorithm}

In Figure~\ref{fig:bcut_bnn_density}, we highlight the distinctions between the standard bispectrum with a fixed cutoff radius, denoted as $\B_{\textrm{cut}}$, and the proposed approach that utilizes a fixed number of neighbors, denoted as $\B^\mathrm{Heavyside}_{nn}$ or $\B^\mathrm{tanh}_{nn}$ (depending on the weight function employed).
We set up NPT simulations of BCC Fe at 100 K, during which the pressure was gradually increased to around 40 GPa (corresponding to a volumetric compression of approximately 20\%). The variations of the $\B_{cut}$ and $\B_{nn}$ components with respect to the volumetric ratio $V/V_0$ are illustrated in Figure~\ref{fig:bcut_bnn_density}-a, b, c, respectively.

The two ways of computing the bispectrum lead to very different results. Indeed, $\B_{nn}$ is almost insensitive to density while the $\B_{cut}$ components evolve significantly with it. Thus, the $\B_{nn}$ should be better at identifying crystal structures even during MD simulations involving a large change in material density. Besides, one can notice a difference between $\B^\mathrm{Heavyside}_{nn}$ and $\B^\mathrm{tanh}_{nn}$. Indeed, even if it does not depend on local density, $\B^\mathrm{Heavyside}_{nn}$ displays some slight evolution that is caused by thermal fluctuations around the sharp step of the Heavyside function. On the contrary, $\B^\mathrm{tanh}_{nn}$ appears satisfyingly stable with density, exhibiting minor sensitivity. 
In the following, we consider BSO4 descriptors computed using either a fixed cutoff radius or a fixed target number of neighbors using the $\tanh$ regularization. The two descriptors will be respectively labelled $\B^N_{nn}$ and $\B^R_{\mathrm{cut}}$ with $N$ and $R$ the values of the corresponding target number of neighbors or cutoff radius.

\subsubsection{Size of the database}
As described above, configurations included in the database are selected at different temperatures and after distinct instantaneous deformations. The combination of these two effects makes the local environment of each atom of the supercell unique. Hence, there is no need to apply a peculiar sparsification procedure to avoid redundancy in the database.

For each of the four different crystal structures considered, 100 snapshots containing 864 atoms each are extracted from the NPT trajectory up to 2/3 $T_m$. Then, the 6 non-zero components of the deformation gradient tensor $\TT{F}$ are sampled using the Latin Hypercube Sampling with Multi-Dimensional Uniformity (LHSMDU)~\cite{lhsmdu}, in order to obtain 100 draws to be applied to the different snapshots, while ensuring the condition $\epsilon_{eq}<0.05$. Following this procedure, our total database contains $N_\mathrm{atoms}$ x $N_\mathrm{snapshots}$ = 864 x 100 = 86400 bispectrum vectors $\mathbf{B} \in \mathbb{R}^{55}$, for each crystal structures, giving a total of $M=345600$ $\mathbf{B}$ vectors. 

\section{Crystal Structure Classifier }
\label{sec:training_validation}
In this section we present the different steps of our procedure to build the supervised learning crystal structure analysis (SL-CSA) tool.
Firstly, we delve into the configuration of our current classifier, which is constructed through a two-step process involving dimensionality reduction and logistic regression (LR).
Subsequently, we compare our classification models to established tools in the literature, with particular emphasis on the adaptive common neighbor analysis (a-CNA), polyhedral template matching (PTM) and diamond structure identification (IDS), three methods available in \texttt{OVITO}~\cite{ovito_ref}.

\subsection{Dimensionality reduction}
\label{subsec:dimensinality}
The database is composed of 4 different crystal structures namely body centered cubic (BCC), face-centered cubic (FCC), hexagonal close packed (HCP) and cubic diamond (c-DIA), each containing 86400 local atomic environments encoded by descriptor vectors $\B \in \mathbb{R}^{55}$.
For the initial step of classification, we perfomed a supervised dimensionality reduction.
We have uses the linear discriminant analysis (LDA), a statistical technique that is commonly used for supervised classification and feature extraction in machine learning~\cite{fisher_1936}. LDA is particularly relevant for dimension reduction while preserving the most important discriminatory information.
The underlying assumption is that the covariance matrix of each class is the same.
LDA works by finding a linear combination of the original features that maximizes the separation between different classes in the data.  The separation between classes is achieved by maximizing the ratio of the between-class variance to the within-class variance. The resulting linear combination, or discriminant function, is then used to project the data onto a lower-dimensional space, with dimension equal to the number of labels reduced by one. In the general case, one can reduce the dimension of the initial descriptor $\B \in \mathbb{R}^D$ leading to a new projected descriptor $\mathbf{x} = P_{\mathrm{LDA}}(\B)$ $: \mathbb{R}^{D} \to \mathbb{R}^{d=C-1}$:
\begin{equation}
    P_{\textrm{LDA}}(\B) = \mathbf{C}_\mathrm{LDA}^\mathrm{T}\cdot (\B-\mub^{\B}_\mathrm{db})
\end{equation}
with $\mathbf{C}_\mathrm{LDA} \in \mathbb{R}^{Dxd}$ the reduction coefficients matrix of LDA and $\mub^{\B}_\mathrm{db} \in \mathbb{R}^{D}$ the average descriptor of the entire database. In the present case, the initial dimension $D=55$ corresponds to the BSO4 dimension imposed by the choice $J_\mathrm{max}=4$ and the new projector has a dimension $d=3$ equal to the number of crystal structures of the database minus one. Thanks to LDA, the separation between classes is maximized in this low dimensional space, hence facilitating the subsequent classification step.

\subsection{Logistic Regression}
\label{subsec:logistic_regression}
Once the dimension reduction step is performed by means of LDA, the new descriptor $\mathbf{x}$ is considered as an input for performing a multinomial logistic regression which provides a probability vector $\mathbf{p} = P_{\textrm{LR}} (\mathbf{x})$ $P: \mathbb{R}^{d} \to \mathbb{R}^C$ defined as:
\begin{equation}
    P_{\textrm{LR}} (\mathbf{x}) = \frac{\exp (\mathbf{s}(\mathbf{x}))}{\sum_i \exp (\mathbf{s(\mathbf{x})}[i] )}
\end{equation}
where $\mathbf{s}(\mathbf{x}) \in \mathbb{R}^{C}$ corresponds to the score vector that reads:
\begin{equation}
    \mathbf{s}(\mathbf{x}) = \mathbf{b}_\mathrm{LR} + \mathbf{D_\mathrm{LR}} \cdot \mathbf{x}^\mathrm{T}
\end{equation}
with $\mathbf{b}_\mathrm{LR} \in \mathbb{R}^{C}$ and $\mathbf{D}_\mathrm{LR} \in \mathbb{R}^{Cxd}$ the bias vector and decision matrix of the logistic regression model after training. In the end, the crystal structure assigned to each atom with descriptor $\mathbf{x}$ is computed as the \textit{argmax} of the probability vector $\mathbf{p}(\mathbf{x})$. In the present work, $C=4$ e.g. the total number of crystal structures. The logistic regression step will systematically attributes a crystal structure to an atom which can lead to misclassification. Some atoms, e.g. defective ones, will be wrongly classified as crystalline. 
This misclassification is expected because the LDA dimension-reduced descriptors $\mathbf{x}$ are constructed based on the assumption that the covariance matrix of each class is identical.
This pitfall can be overcome by methods such as QDA (Quadratic Discriminant Analysis), which are less strict and allow for different feature covariance matrices for different classes. However, these methods result in a quadratic decision boundary, which is more challenging to train and stabilize.
For this reason, we stick to the framework of LDA and make the final decision in the classification by employing statistical distances with respect to each class based on the full covariance matrix of each class, similar to \cite{Goryaeva2020, goryaeva_compact_2023}.
Consequently, an additional step, serving as a sanity check (referred to as step 4 in Algorithm~\ref{alg:classifier}), is performed and described in the subsequent section.

\subsection{Crystal structure classifier}
\label{subsec:classifier_definition}
\begin{figure}[h!]
    \centering
    \includegraphics[width=\linewidth]{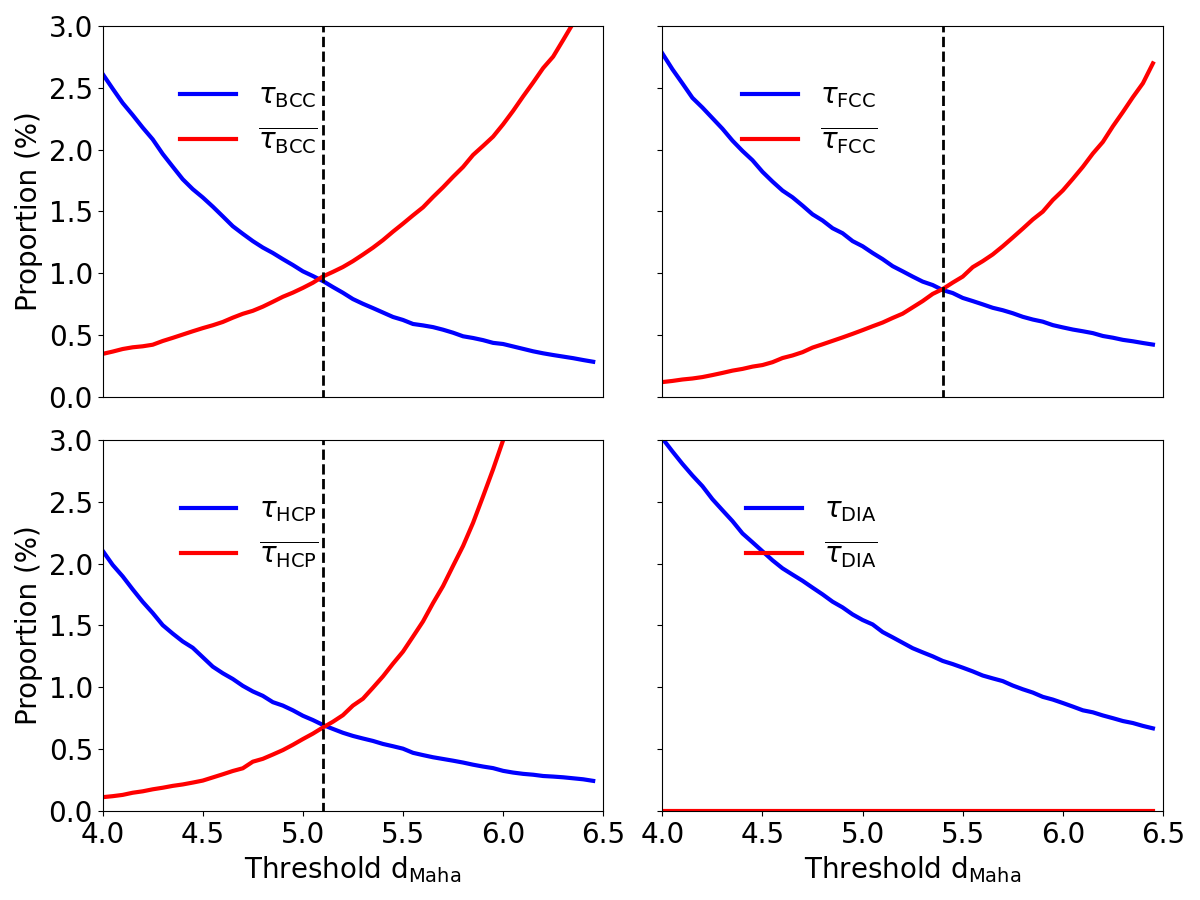}
    \caption{Evolution of the error rates as a function of the acceptance threshold for each crystal structure.}
    \label{fig:thresholds}
\end{figure}

The full database presented in \ref{sec:database} is replicated into six different databases computed with different flavors of the descriptors. We use the $\B^R_\mathrm{cut}$ descriptor with $R$ equal to 3.0, 6.0, and \SI{9.0}{\angstrom}, and the $\B^N_\mathrm{nn}$ descriptor with $N$ set to 24, 48, and 72 respectively. A different instance of our classification model (SL-CSA) is trained on each of the six databases. The results obtained with each model are compared against each other and against crystal structure classifiers of the literature in the next section.
\begin{figure}[t!]
    \centering
    \includegraphics[width=\linewidth]{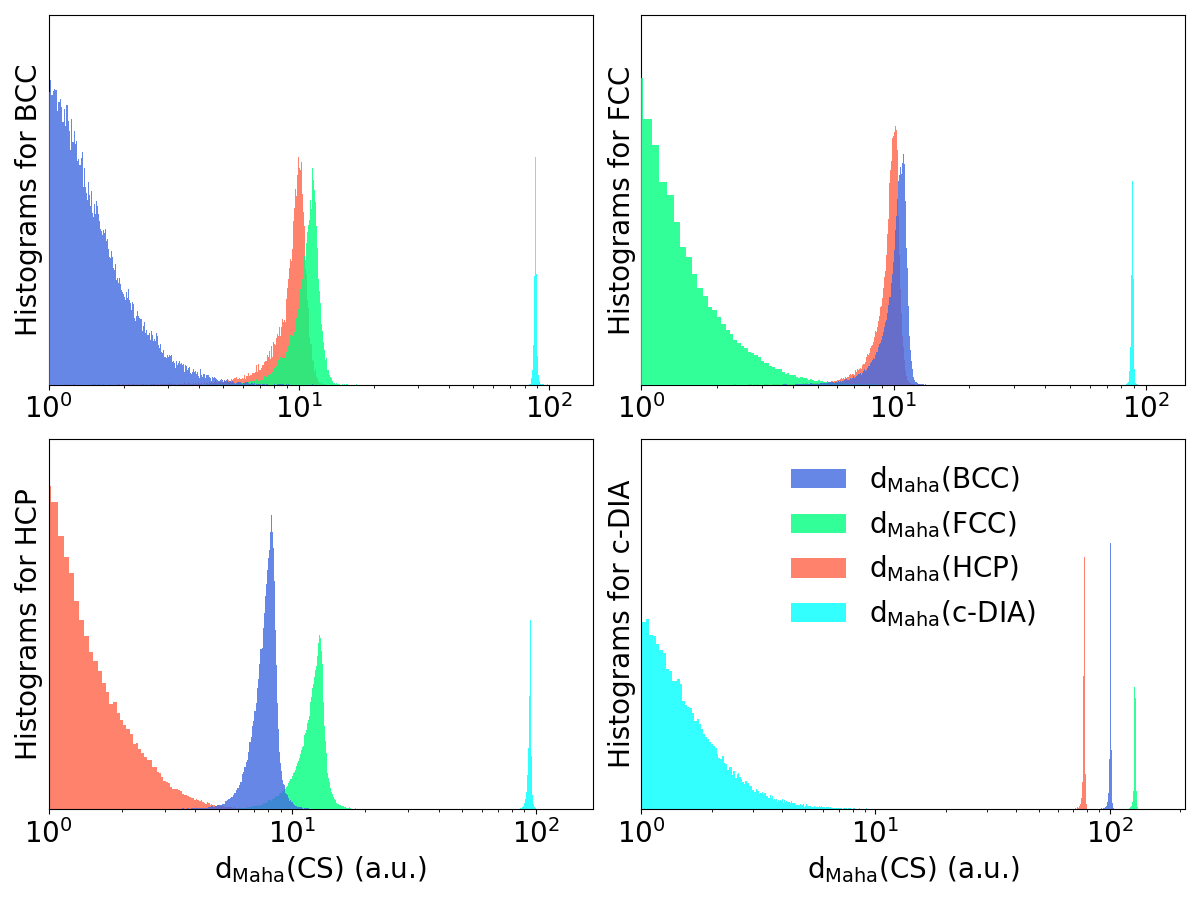}
    \caption{Histograms of between- and within-class Mahalanobis distance for each crystal structure of the entire database. The present case correspond to the database with descriptor $\B^{24}_\mathrm{nn}$.}
    \label{fig:maha_threshold}
\end{figure}
The purpose of the classification procedure is to identify local crystal structure with high fidelity or to detect atoms that do not belong to any of the reference crystal structures, considered as outliers. Since the logistic regression assigns a crystal structure to all atoms, we consider a final sanity check step using the Mahalanobis distance of a given recued descriptor $\mathbf{x}$ to a given crystal structure cs:
\begin{equation}
d^\mathrm{cs}_\mathrm{Maha} (\mathbf{x}) = \sqrt{ (\mathbf{x} - \mub_\mathrm{cs} )^\mathrm{T} \cdot \TT{\Sigma}^{-1}_\mathrm{cs} \cdot (\mathbf{x}-\mub_\mathrm{cs})}
\end{equation}
where $\TT{\Sigma}_\mathrm{cs}$ and $\mub_\mathrm{cs}$ are the sample covariance matrix and average of the descriptors in the reference database associated to crystal structure CS. The decision to assign the cs to the descriptor $\T{x}$ is made based on a threshold criterion on this distance: if the distance is lower than an acceptance threshold, the cs is assigned to the descriptor. Hence, this acceptance threshold determines the accuracy of the classifier. It is chosen by investigating the properties of the distributions of distances in the reference database: for each descriptor, the distance to each cs is computed, and the results are gathered by crystal structure so that the distribution of within- and between-class distances can be computed for each crystal structure. Results corresponding to $\B^{24}_{nn}$ are shown in Figure~\ref{fig:maha_threshold} whereas the distributions for other descriptors are given in the Supporting information (see Figures S1 to S4).
The separation between c-DIA and other classes is systematically large due to the geometrical particularities of the diamond structure compared to BCC, FCC and HCP. On the other hand, between-class distances for BCC, FCC and HCP crystal structures are smaller and may even overlap. The acceptance threshold needs to be carefully chosen to ensure a minimal error during classification.
In order to calculate this threshold for each class, we define the error rate of a crystal structure cs as:
\begin{equation}
    \tau_\mathrm{cs} = \frac{n_{d^\mathrm{cs}_\mathrm{Maha} \geq \mathrm{threshold}}}{n^\mathrm{cs}_\mathrm{tot}}
\end{equation}
where $n_{d^\mathrm{cs}_\mathrm{Maha}}$ corresponds to the number of atoms of the database with a distance greater than the threshold and $n^\mathrm{cs}_\mathrm{tot}$ is the total number of atoms with crystal structure cs. 
We also define a second error rate that quantifies the number of misassigned atoms (i.e. atoms of another crystal structure with a between-class distance with cs lower than the defined threshold):
\begin{equation}
     \overline{\tau_\mathrm{cs}} = \frac{n_{d^\mathrm{\overline{cs}}_\mathrm{Maha} \leq \mathrm{threshold}}}{n_\mathrm{tot}-n^\mathrm{cs}_\mathrm{tot}}
\end{equation}
where $n_{d^\mathrm{\overline{cs}}_\mathrm{Maha}}$ corresponds to the number of atoms belonging to another crystal structure with a distance to cs lower than the threshold, and $n_\mathrm{tot}$ is the total number of samples in the database. Finally, the optimal threshold for each crystal structure is defined as the Mahalanobis distance that minimizes $|\tau_\mathrm{cs}-\overline{\tau_\mathrm{cs}}|$. Both error rates are displayed in blue and red lines in Figure~\ref{fig:thresholds} for each crystal structure as a function of the acceptance threshold. The optimal threshold is displayed as a vertical dashed line and is defined by the crossing of the two error rates.
Since the distributions of Mahalanobis distance are entirely disjointed for the c-DIA case, the acceptance threshold is arbitrarily set to \SI{6}{\angstrom}. This allows to get 99 \% of correct predictions on the database.
For BCC, FCC and HCP crystal structures, the calculated optimal thresholds are equal to 5.1, 5.4 and \SI{5.1}{\angstrom} respectively. This means that atoms exhibiting within-class distances greater than the acceptance threshold for each crystal structure will be considered as defective. 
In addition, the Mahalanobis distance to a specific crystal structure can also be used as a tool to classify defects, as was done previously~\cite{Goryaeva2020} and this will be discussed in Section~\ref{sec:applications}.

\subsection{Comparison to standard classifiers}
\label{subec:comparison}
In the following, we focus on the two classifiers built on the $\B^{24}_\mathrm{nn}$ and $\B^{6}_\mathrm{cut}$, descriptors as they systematically provide better scores when identifying the structures in the testing database. 
The selected classifiers are tested against well-established crystal structure identification methods, namely the a-CNA, PTM, and IDS algorithms (where applicable).
It is important to mention that certain algorithms may require specific configurations. Therefore, in this section, we provide a detailed description of the parameters used in the current study.
Prior to analyzing a particle neighborhood, a-CNA determines the optimal cutoff radius automatically for each individual particle by computing a local length scale specific to each crystal structure. 
It is to be noted that a-CNA does not have a tolerance criterion associated with its classification, i.e. 
if an atom neighborhood cannot be mapped to one of the known crystal structures, it is classified as non-crystalline, and labeled ``unknown". In contrast to a-CNA, PTM requires a user-defined RMSD parameter, typically set to 0.1 by default. This parameter is the same for all structures, and atoms with RMSD greater than the threshold are assigned as non-crystalline. Setting a much larger RSMD threshold can reduce the number of ``unknown" labels, however, it also favors the appearance of false positives.
Calibration of the RMSD parameter for different crystal structures is out of the scope of the present study and we only perform the analysis with the most commonly used standard settings, i.e. RMSD=0.1.
The acceptance threshold of the SL-CSA classifier is defined for each type of structure as described in Section~\ref{subsec:classifier_definition}. 

Below we explore the performance of the structure identification methods for different thermo-mechanical states. In Section~\ref{subsubsec:lowP_highT} we investigate the effect of thermal fluctuations; in Section~\ref{subsubsec:sensitivity_density} sensitivity to the material's density is examined; and finally, in Section~\ref{subsubsec:sensitivity_deformation} we consider the sensitivity to large non-hydrostatic deformation. 

\subsubsection{Sensitivity to high temperature}
\label{subsubsec:lowP_highT}
Here we perform an analysis of NPT trajectories in BCC Iron, FCC Aluminium, HCP Zirconium, and c-DIA Silicon, where pressure was maintained at 0 GPa as the temperature is increased up to each material's melting point $T_m$.
These simulations are distinct from those of the training database.
\begin{figure}[b!]
    \centering
    \includegraphics[width=\linewidth]{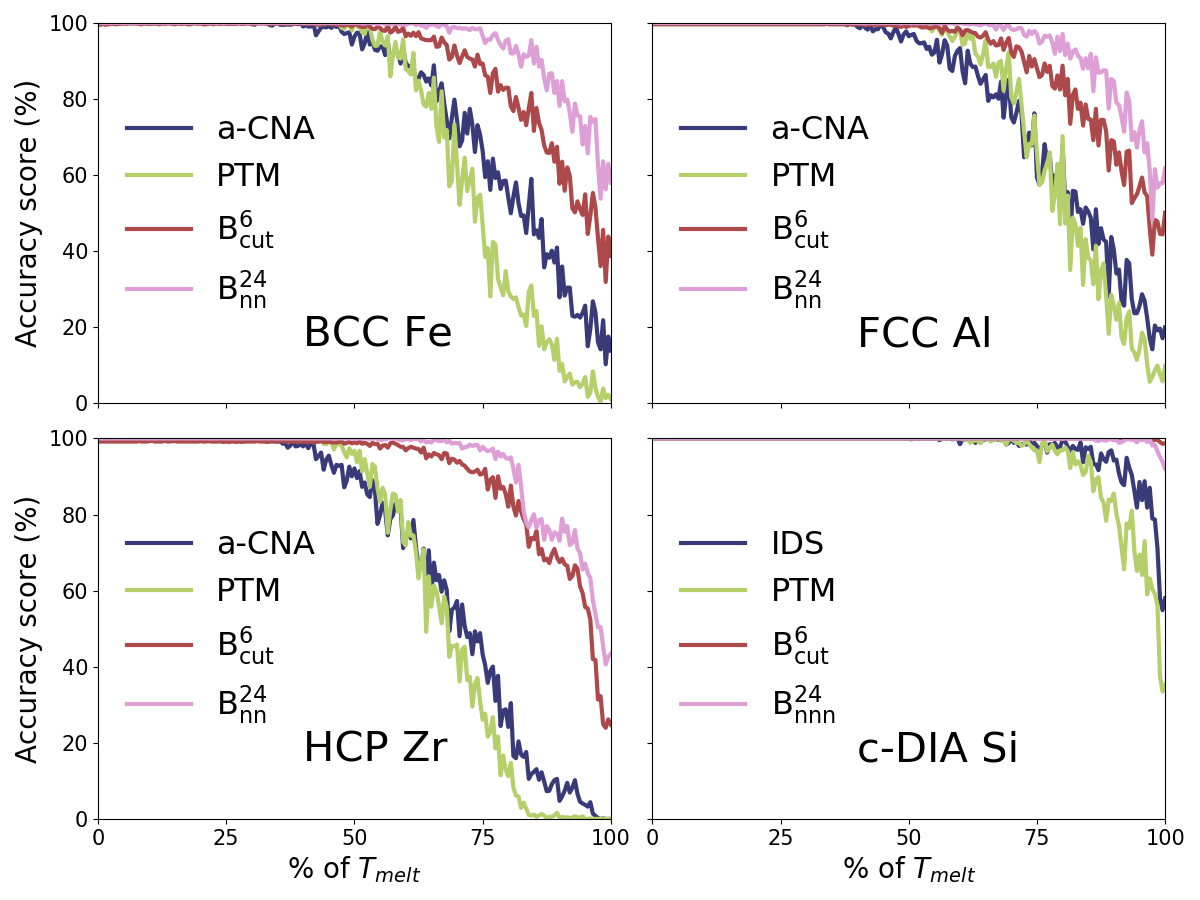}
    \caption{Comparison of the supervised learning crystal structure classifiers with descriptors $\B^6_\mathrm{cut}$ (brown) and $\B^{24}_\mathrm{nn}$ (pink) to a-CNA (blue), PTM (green) and IDS (blue) from \texttt{OVITO}. Each panel corresponds to a NPT trajectory on a different crystal structure, namely BCC Fe, FCC Al, HCP Zr and c-DIA Si.}
    \label{fig:compare_high_T}
\end{figure}
We define the accuracy score of a crystal structure classifier algorithm as the number of atoms identified as crystalline over the total number of atoms in the simulation, assuming that the analyzed materials are fully crystalline at $\frac{2}{3}T_m$. Figure \ref{fig:compare_high_T} reports the evolution of the accuracy score as a function of the average temperature in the simulation cell for the four different crystal structures. 
The SL-CSA classifier clearly outperforms a-CNA, and PTM for BCC Fe, FCC Al, and HCP Zr. The conventional methods shift from 100 \% accuracy before reaching $\frac{2}{3}T_m$, while the SL-CSA retains more than 98 \% accuracy along the whole NPT trajectory. The IDS tool used for c-DIA Si appears not very sensitive to thermal noise and provides comparable results with SL-CSA with an accuracy above 99 \% for this structure along the entire trajectory. 
 Finally, the classifier built with $\B^{24}_\mathrm{nn}$ appears less sensitive to thermal fluctuations than its counterpart using $\B^{6}_\mathrm{cut}$. 

\subsubsection{Sensitivity to material's density}
\label{subsubsec:sensitivity_density}
In order to examine the performance of our classifiers in extrapolation conditions with changing density (beyond their trained domain at a density of $\rho_0$), we generate 100 synthetic samples for each of the four crystal structures at densities $\rho \in [0.5\rho_0, 1.5\rho_0]$. 
These samples correspond to simulation cells at different lattice parameters containing 864 atoms, each atom being shifted from its crystalline position by Gaussian noise with $\sigma=0.1$ \AA. This setup allows building artificial configurations far from the domain of validity of the interatomic potential, spanning a wide range of material densities. 
\begin{figure}[!h]
    \centering
    \includegraphics[width=\linewidth]{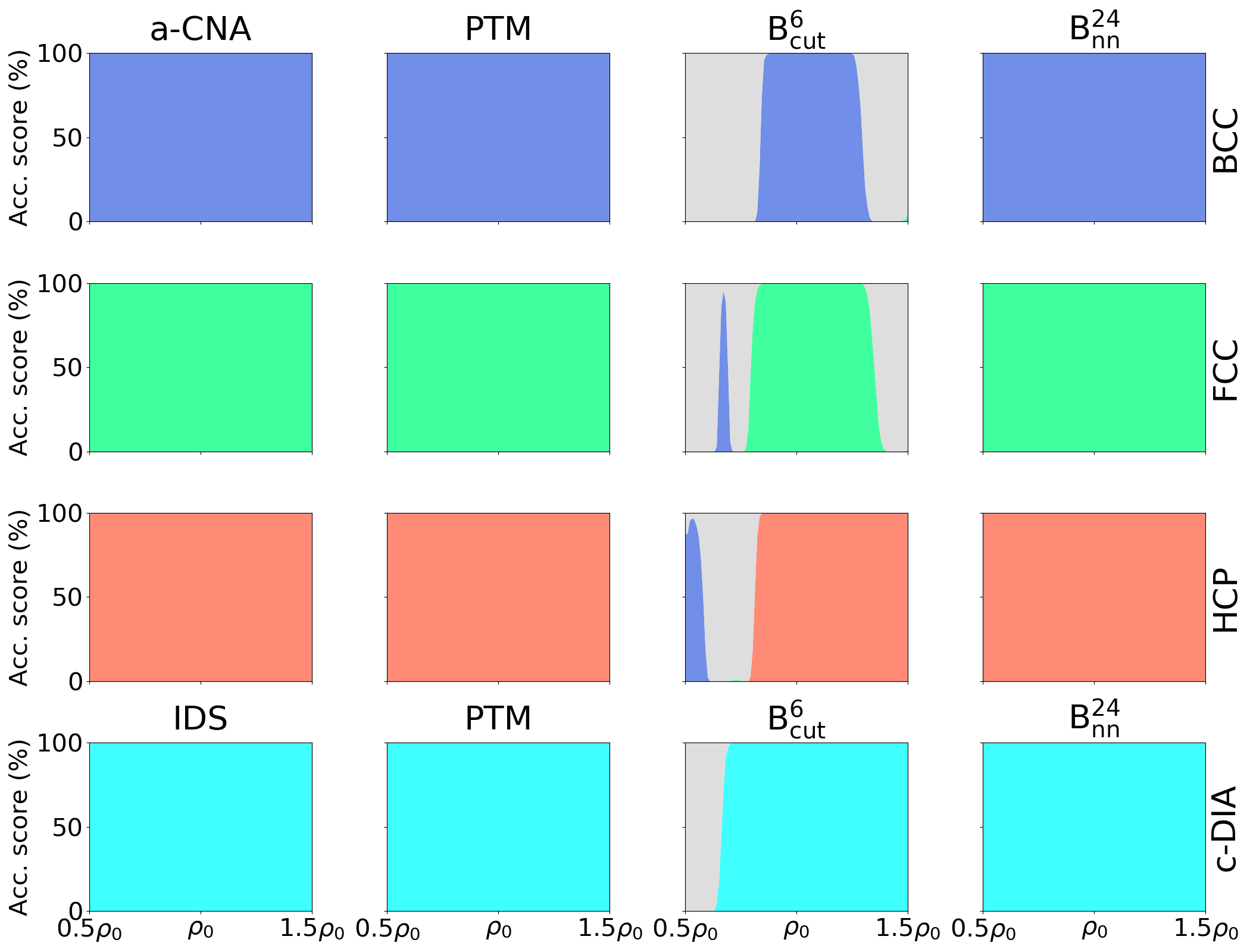}
    \caption{Comparison between classifiers for the prediction of crystal structures as a function of material density. Each row concerns a different crystal structure, while each column is associated with a different classifier, namely a-CNA (or IDS for c-DIA Si), PTM, and the two supervised learning presented in this work.}
    \label{fig:compare_rho}
\end{figure}

Figure~\ref{fig:compare_rho} compares the results between a-CNA, IDS, PTM, and the two classifiers based on $\B^{6}_\mathrm{cut}$ and $\B^{24}_\mathrm{nn}$ descriptors. 
Here, $\rho_0$ is taken as the reference density for all measurements and classifiers since we aim at comparing the results w.r.t. the samples present in the database. For the three metallic systems, 
a-CNA and PTM algorithms provide very good results and correctly predict 100 \% of the crystal structures in the entire range of density spanned. 
The CNA-based classifiers are not adapted for the diamond structure, and we use the IDS classifier instead. Both IDS and PTM predict 100\% of c-DIA structure for any density, proving that they are not sensitive to volumetric strain. 
Thus, the three tested standard classifiers can perform well in the presence of large volumetric deformation, at least in the presence of moderate thermal noise.
The $\B^{6}_\mathrm{cut}$ based classifier performs well for the densities near $\rho_0$. However, due to the fixed cutoff, it largely fails at predicting correct crystal structures when density changes and may even predict a wrong crystal structure, e.g. BCC instead of FCC/HCP.
Using the classifier trained on the $\B^{24}_\mathrm{nn}$ descriptor allows for the full restitution of the correct crystal structures in BCC, FCC, HCP, and c-IDA systems, over the whole range of densities. Together with the low sensitivity to thermal noise, these results prove its applicability to various thermodynamic conditions, i.e. when large hydrostatic pressure and high temperature are involved. 

\subsubsection{Sensitivity to large deformation}
\label{subsubsec:sensitivity_deformation}
In order to explore the sensitivity of the classifiers to large deformations, we design a database with NPT trajectories for the four tested structures. For each material, the trajectories are performed at ambient pressure and at $\frac{2}{3}T_{m}$  for 200 ps. Along each trajectory, 200 snapshots are extracted for subsequent application of large deformations. All configurations are different from those of the learning database, and simulations were carried out with different seeds for temperature initialization. 

In order to explore independently diagonal and deviatoric deformations, we consider two subsets of structures, where ($F_{11}$, $F_{33}$) and ($F_{12}$, $F_{13}$) are applied.
Each component of the deformation tensor $F_{ij}$ is drawn from a uniform distribution in the intervals [0.7, 1.3] and [-0.3, 0.3] for longitudinal and deviatoric strains components respectively.
\begin{figure*}
    \centering
    \includegraphics[width=0.49\linewidth]{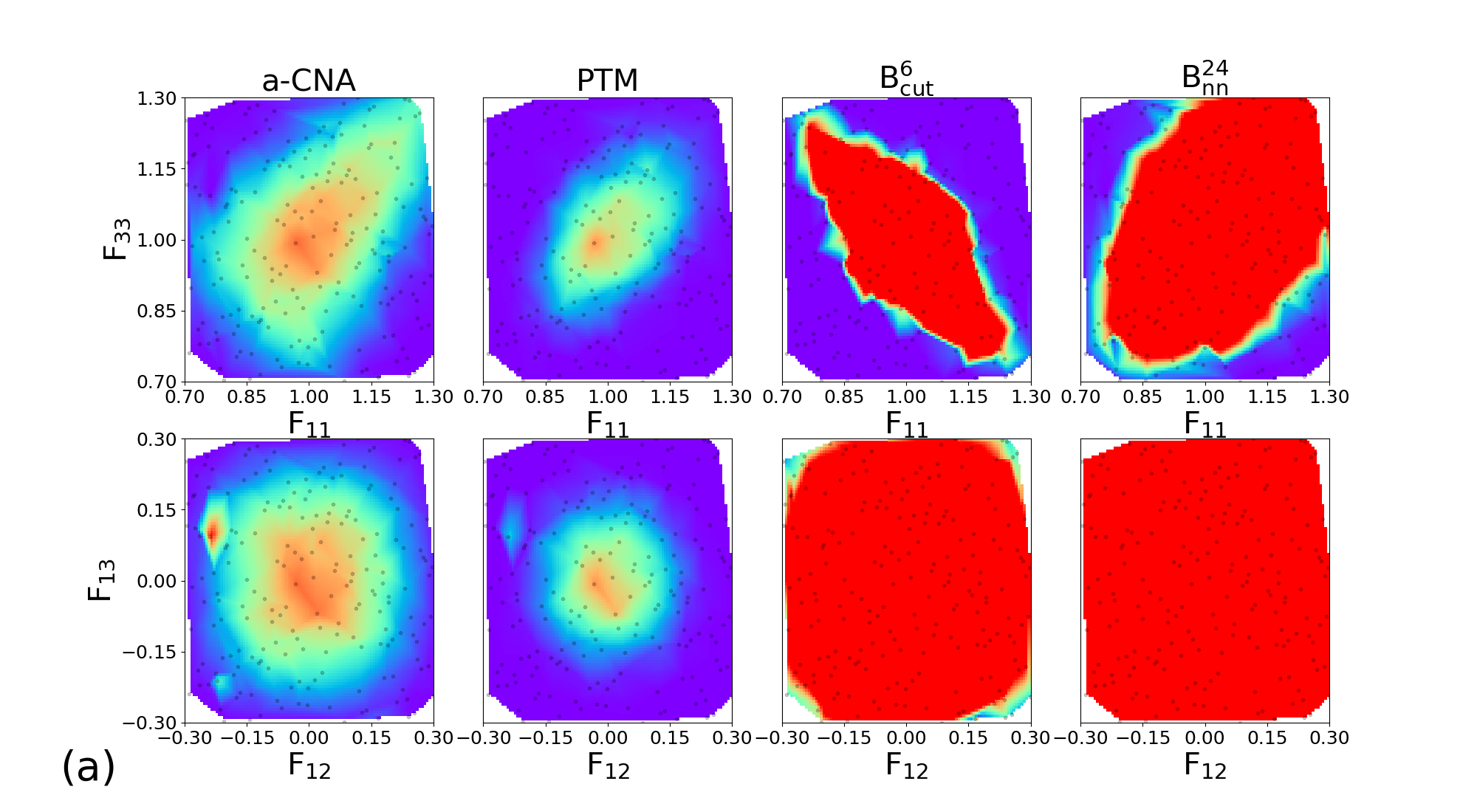}
    \includegraphics[width=0.49\linewidth]{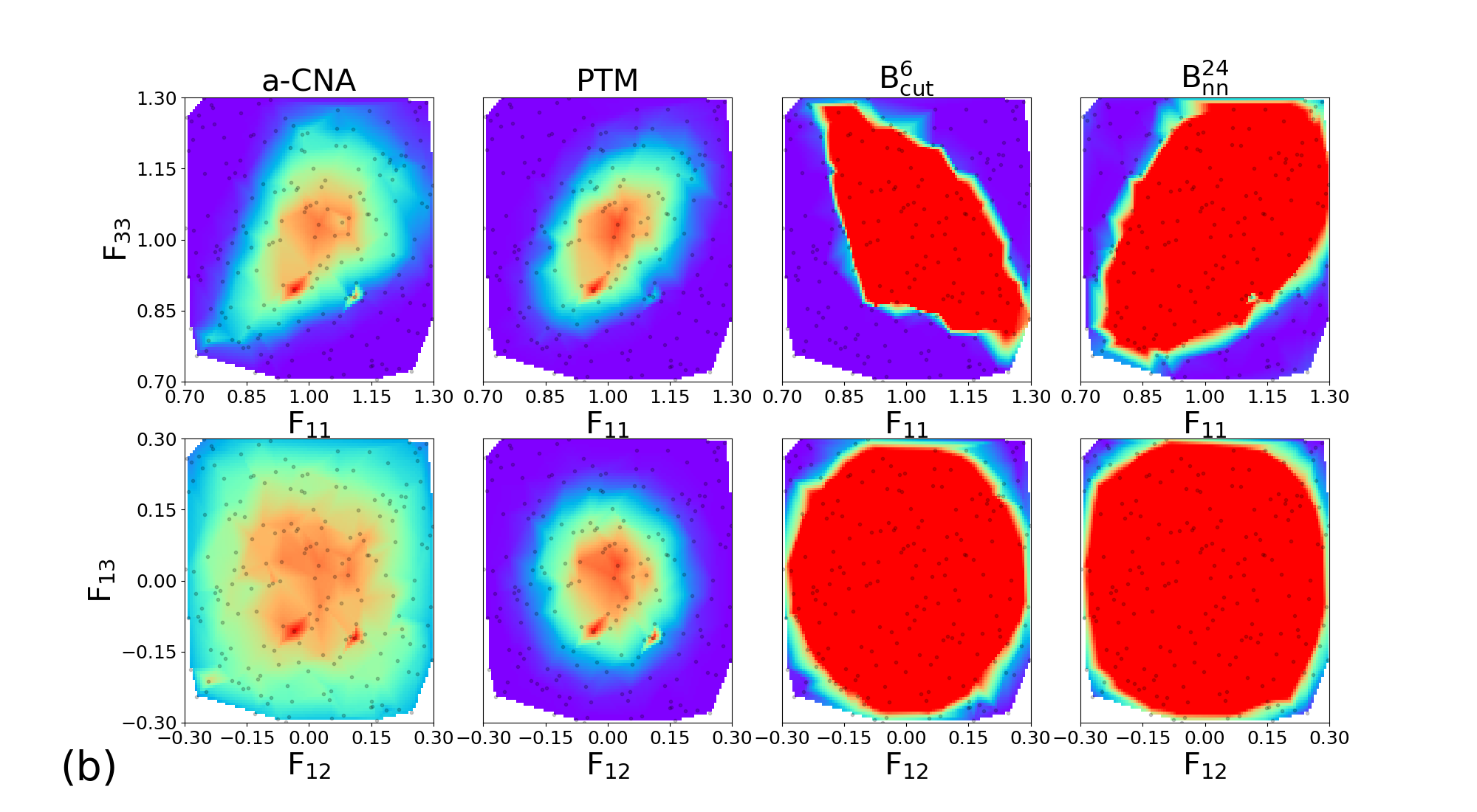} \\
    \includegraphics[width=0.49\linewidth]{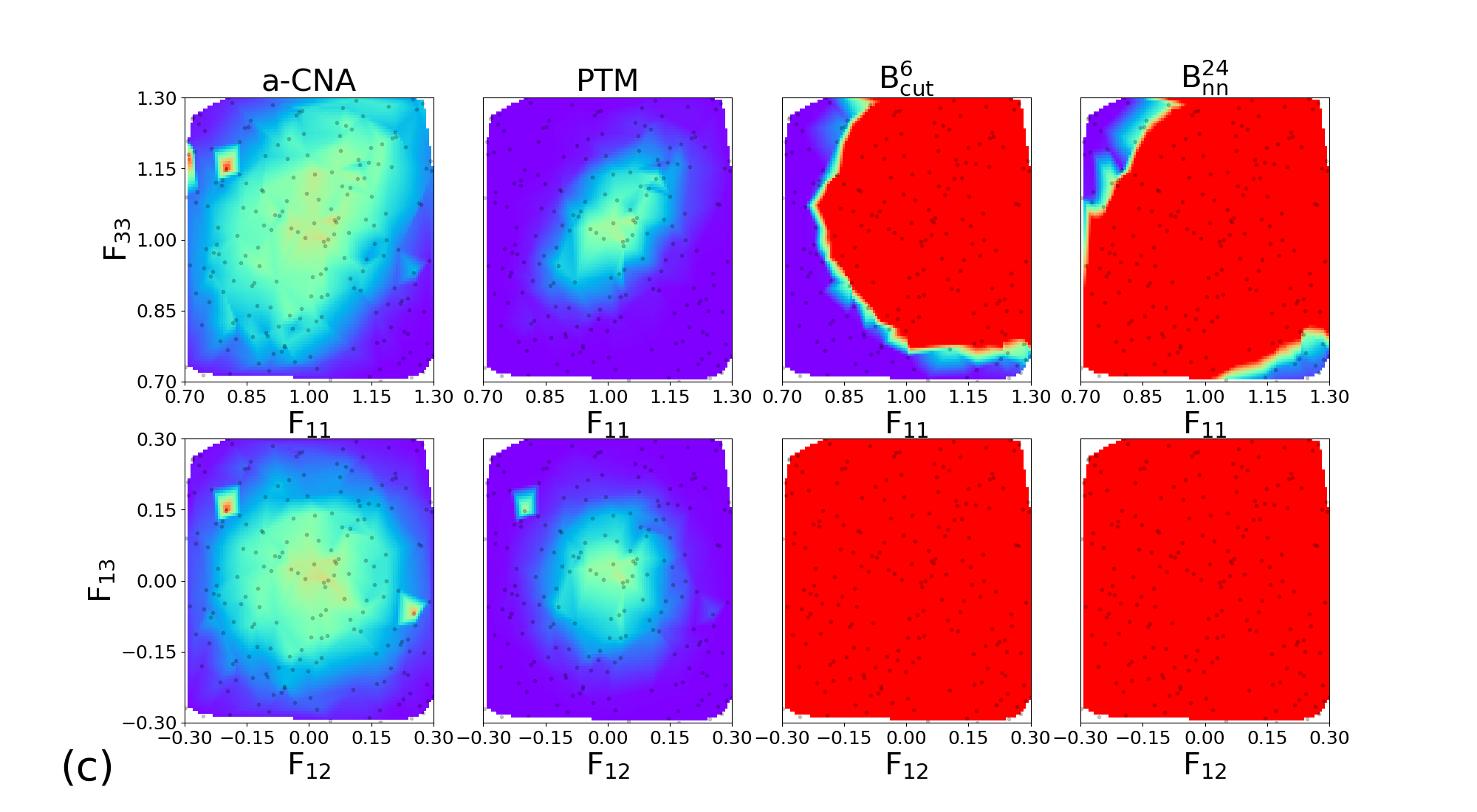}
    \includegraphics[width=0.49\linewidth]{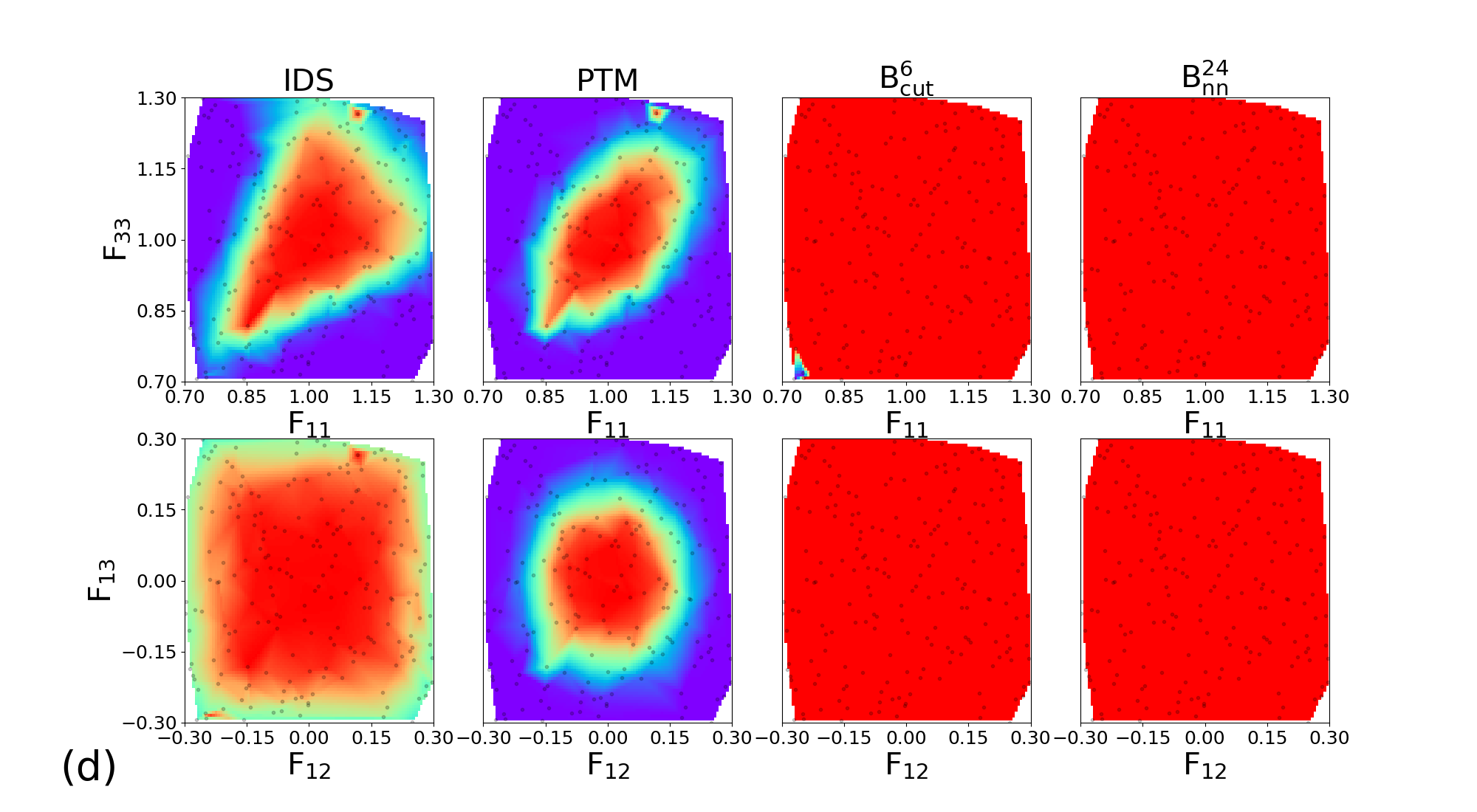} \\    
    \caption{Comparison between classifiers for the prediction of crystal structures as a function of various deformation tensor components. The subfigures \textbf{a, b, c, d} correspond to BCC, FCC, HCP and c-DIA structures, respectively. Each column of each subfigure denotes the employed method, i.e. a-CNA, PTM, $\B_\mathrm{nn}^{24}$ and $\B_\mathrm{cut}^{6}$. On each subfigure, we represent employing a color map the percentage of correctly predicted atomic environments as a function $(F_{11}, F_{33})$ or $(F_{12}, F_{13})$, for the first and second line, respectively.  On each subfigure, the color indicates the accuracy of the corresponding method, going from 0 \% (purple) to 100 \% (red).}
    \label{fig:compare_def}
\end{figure*}
The 200 couples of diagonal deformation tensor components ($F_{11}$, $F_{33}$) are assigned to the 200 snapshots of each crystal structure, by applying the corresponding deformation gradient tensor $\T{F}$ to the simulation cell while rescaling atomic positions. Finally, the different classifiers a-CNA, PTM, $\B_\mathrm{cut}^{6}$ and $\B_\mathrm{nn}^{24}$ are used to analyse the deformed samples. The same process using deviatoric deformation tensor components ($F_{12}$, $F_{13}$) has been employed. Results for each crystal structure are displayed in Figure~\ref{fig:compare_def}.

Similar trends are observed for BCC, FCC, HCP, and c-DIA, where the performance of the classifiers is roughly independent of the crystalline structure. For diagonal deformations associated with compression and tension of the simulation cell, the CNA and PTM analysis are limited to small or very small deformation only (lower than 5\%), and their accuracy is strongly reduced beyond this point. This effect is likely attributed to the combined effects of deformation and temperature. On the other hand, the performances of the $\B_\mathrm{cut}^{6}$ and $\B_\mathrm{nn}^{24}$ classifiers remain robust up to deformation as large as 20\% (30\% in the best cases). We note that $\B_\mathrm{nn}^{24}$ even shows an extended range of accuracy compared to $\B_\mathrm{cut}^{6}$, and both classifiers exhibit some anisotropic response to diagonal deformation.
Concerning deviatoric deformations the performances of our classifiers are even better than CNA and PTM, with correct structural assignment up to 30\% deformation. Once again we attribute the high accuracy of our classifiers to their capabilities to handle temperature and deformation effect conjointly. This is highly valuable in the context of in situ classification of large-scale simulations of materials at extreme conditions. In the following, we demonstrate the applications of our most robust classifier constructed with the $\B^{24}_\mathrm{nn}$ descriptor for the analysis of structures challenging to perform with traditional methods. 

\section{Applications}
\label{sec:applications}
Two examples of interest to the materials science community are outlined below. 
The first example considers the crystalline Zr solid-solid phase transition from HCP to BCC under high pressure, and the second explores the identification of dislocations as they form and expand from a Frank-Read source in Al. 
The difficulty in the first example is in the accurate identification of the crystal structures where volume discontinuity may be present, whereas in the second example it relies on the extraction of defective atoms present in the dislocations' cores. The comparison of the results provided by SL-CSA and traditional methods is made for both applications.

\subsection{HCP-BCC phase transition in HCP Zr}

\begin{figure}[h!]
    \centering
    \includegraphics[width=\linewidth]{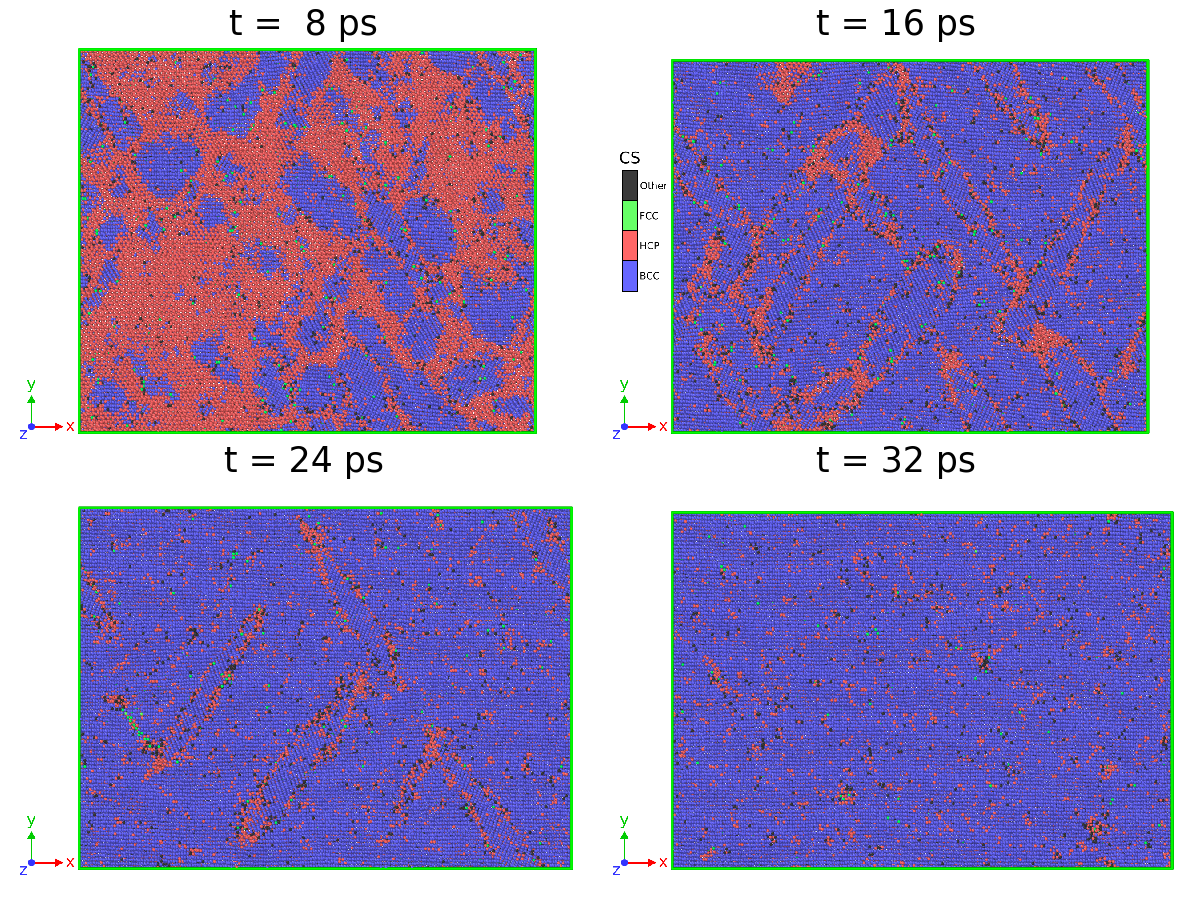} 
    \caption{Four snapshots of the MD simulation of the HCP $\to$ BCC phase transition are shown. These snapshots were taken immediately after the first nucleation event of a BCC seed. The local crystal structure was determined using the SL-CSA classifier, with HCP atoms depicted in red and BCC atoms depicted in blue.}
    \label{fig:snapshots_hcp_bcc}
\end{figure}

Here we perform the analysis of the high-pressure HCP $\to$ BCC transition in Zr, following the MD simulation procedure previously described in Refs. \cite{Willaime1989, Ahuja1993, Greeff2005, Zong2019, Xia1991}.
We aim to reveal the effect of the accuracy of the classification on the characterization of this transition. 
To this end, we track the number of atoms belonging to different crystalline structures as a function of time. 
To investigate this transition, we performed MD simulations in a simulation cell with 442,368 atoms at high temperature and high pressure.
The system was first equilibrated at 0 GPa and 1500 K for 10 ps in the NPT ensemble using a 2 fs timestep. 
Then, the target pressure was set to 18 GPa, allowing the simulation cell to relax independently  in the three directions of space. The temperature was maintained at 1500 K for the entire simulation with a Nosé-Hoover thermostat. At high pressure, a few ps are needed for the first BCC seed to nucleate in the HCP single crystal, meaning that the BCC structure becomes thermodynamically more stable than its HCP counterpart. 
Four snapshots of the MD simulation are taken at different times and are depicted in Figure~\ref{fig:snapshots_hcp_bcc}.
\begin{figure}[h]
    \centering
    \includegraphics[width=\linewidth]{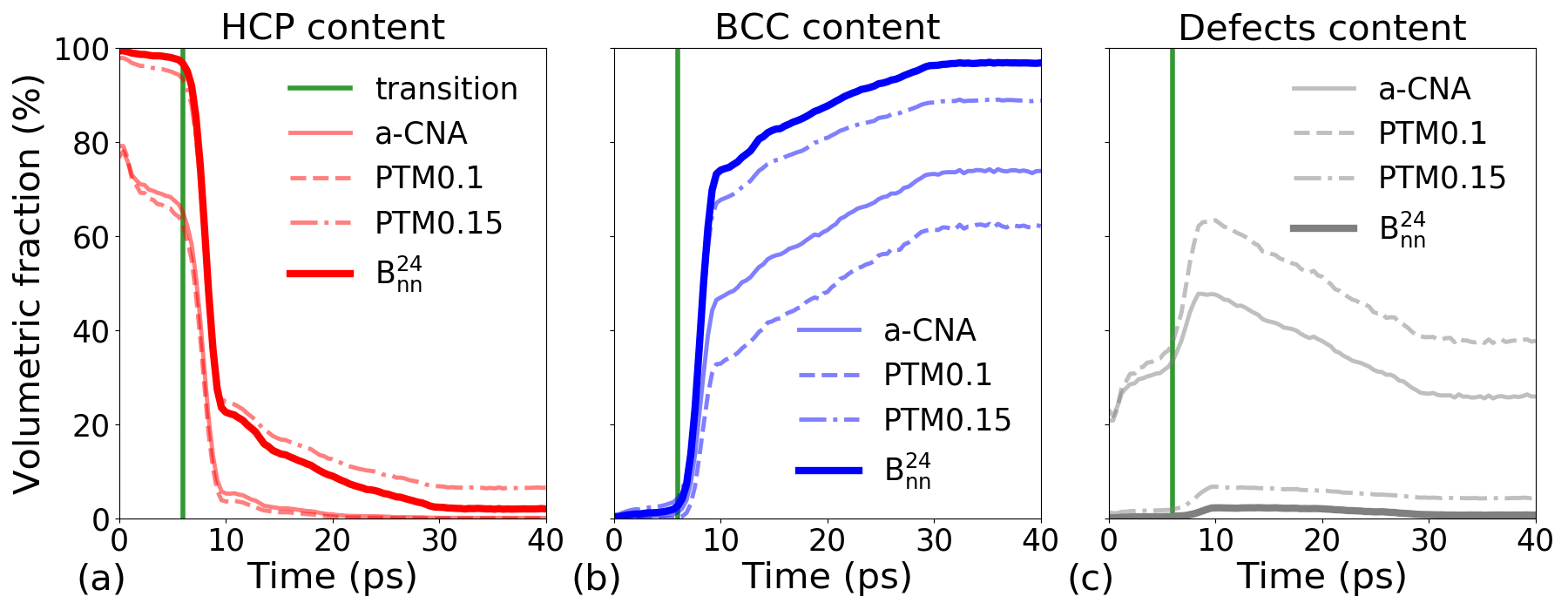}
    \caption{Evolution of BCC \textbf{(a)}, HCP \textbf{(b)} and non-crystalline populations \textbf{(c)} vs time during the HCP to BCC phase transition. Solid lines are for a-CNA, dashed lines for PTM with RMSD=0.1, dotted-dashed lines for PTM with RMSD=0.15 and bold solid lines for SL-CSA. As a guide for the eye, the green vertical line represents the time at which the transition occurs.}
    \label{fig:fractions_hcp_bcc}
\end{figure}

The results of structure identification during the phase transition revealed unexpected differences between the methods, which can be grouped into two distinct categories- CNA and PTM0.1 for one and PTM0.15 and SL-CSA for the other, as shown in Figure~\ref{fig:fractions_hcp_bcc}.
From the beginning of the simulation, CNA and PTM0.1 correctly identify 80\% of the atoms, this proportion rapidly decreasing to 70\%. Then, as the structural transition occurs, the number of unclassified atoms increases drastically up to 60\%, demonstrating the inability of these methods to extract the correct underlying mechanism. The proportion of BCC atoms finally increases slowly, reaching asymptotic values between 60 and 70\%, well below the expected proportion. In this context, using these results to characterize the kinetic of this phase transition would probably lead to wrong predictions. 

On the other hand, PTM0.15 and SL-CSA exhibit similar trends. 
Initially, there is a high population of HCP atoms followed by a rapid transition towards the BCC structure.
Finally, the system reaches an asymptotic behavior, resulting in a significant population of BCC atoms. The population of unclassified atoms during the transition remains low. 
Although these two methods show similar behavior, there are still some differences, especially during the transition phase. SL-CSA exhibits a lower population of unclassified atoms compared to PTM0.15. Additionally, at the end of the simulation, there is an approximate 10\% difference in the population of BCC atoms between the two methods, with SL-CSA leading to a higher population of the BCC phase.
Based on these results, we conclude that the SL-CSA classifier yields better results and, that this method is well suitable for a quantitative evaluation of the mechanism and kinetics of phase transitions. In addition, the proposed method does not require any parameter-tuning in opposition with PTM and its corresponding RMSD, to which the results are very sensitive. 
\begin{figure*}
    \centering
    \includegraphics[width=0.245\linewidth]{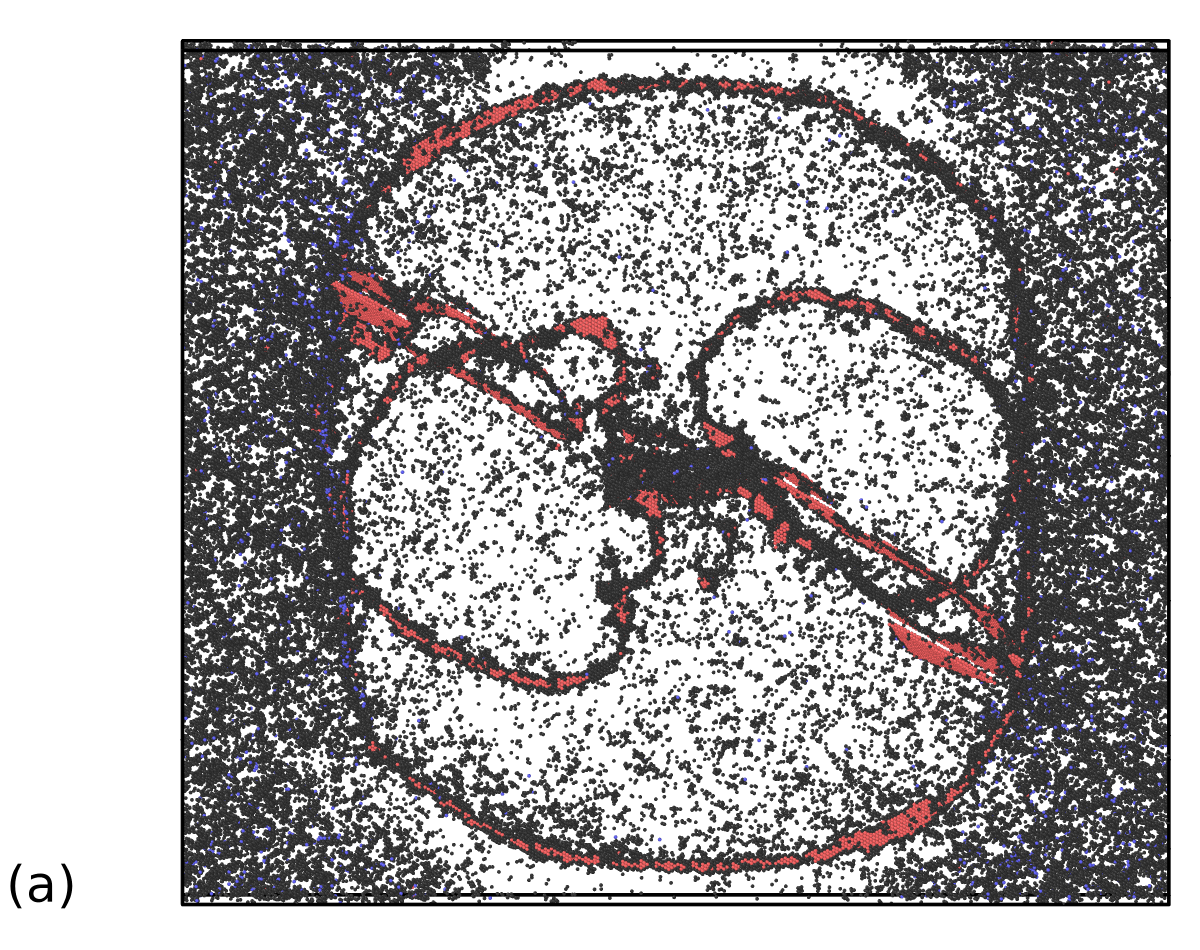}  
    \includegraphics[width=0.245\linewidth]{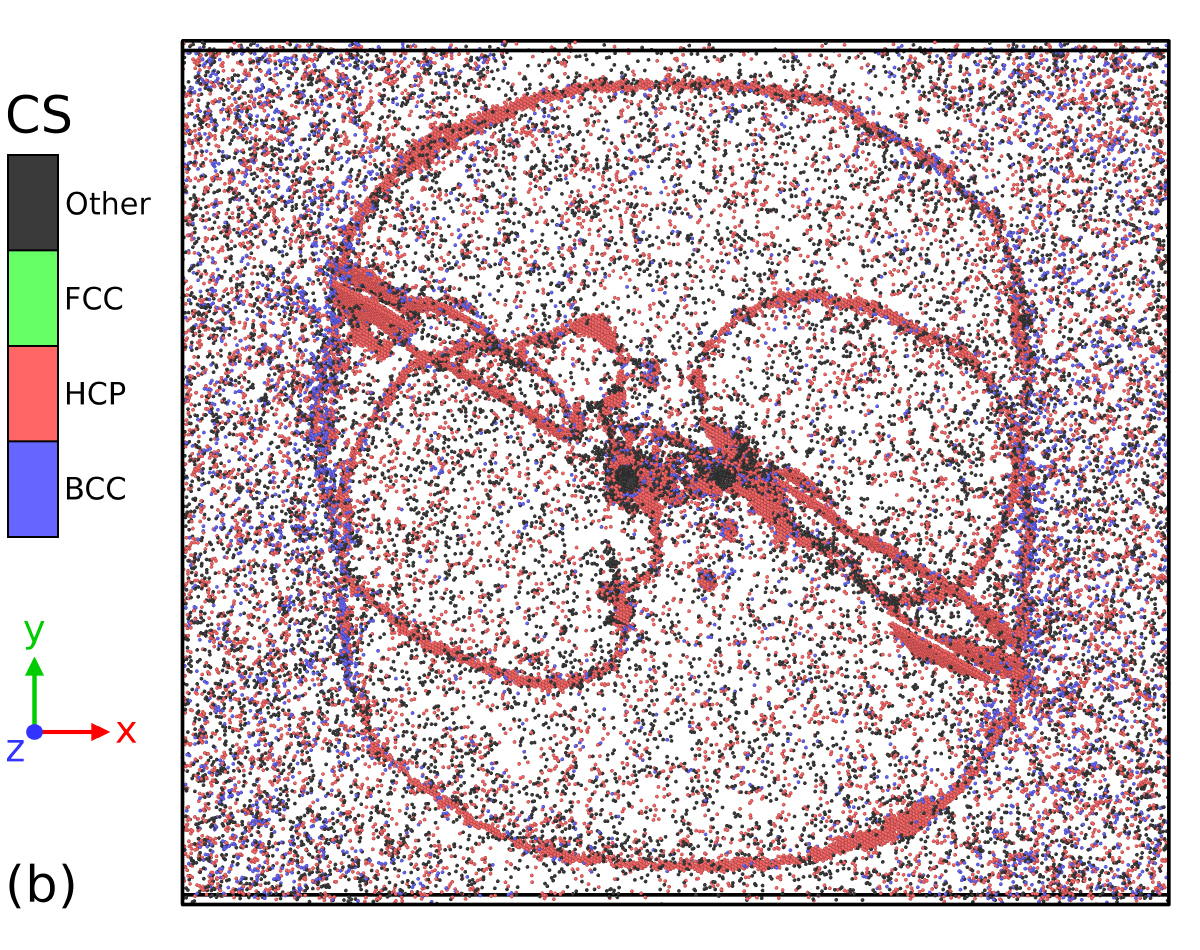}
    \includegraphics[width=0.245\linewidth]{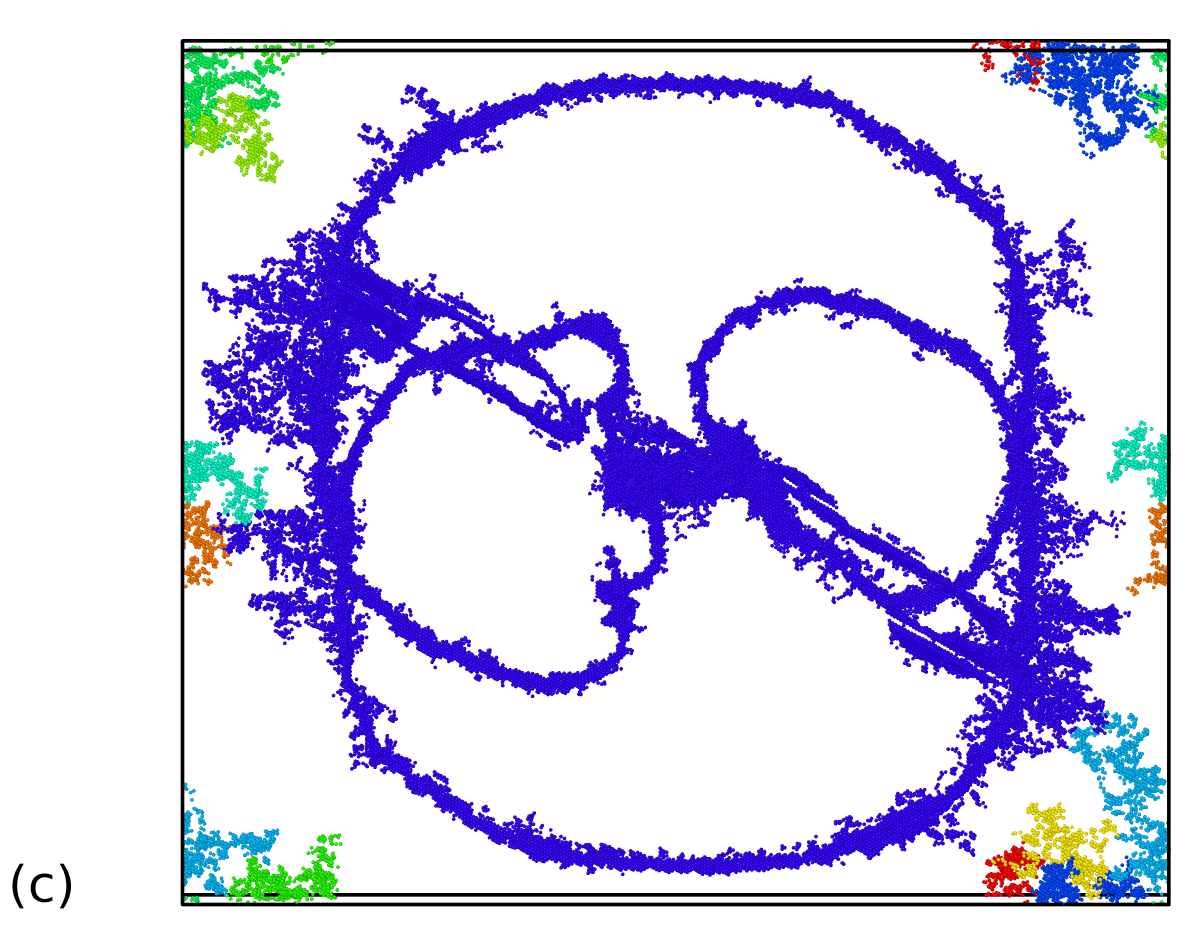}  
    \includegraphics[width=0.245\linewidth]{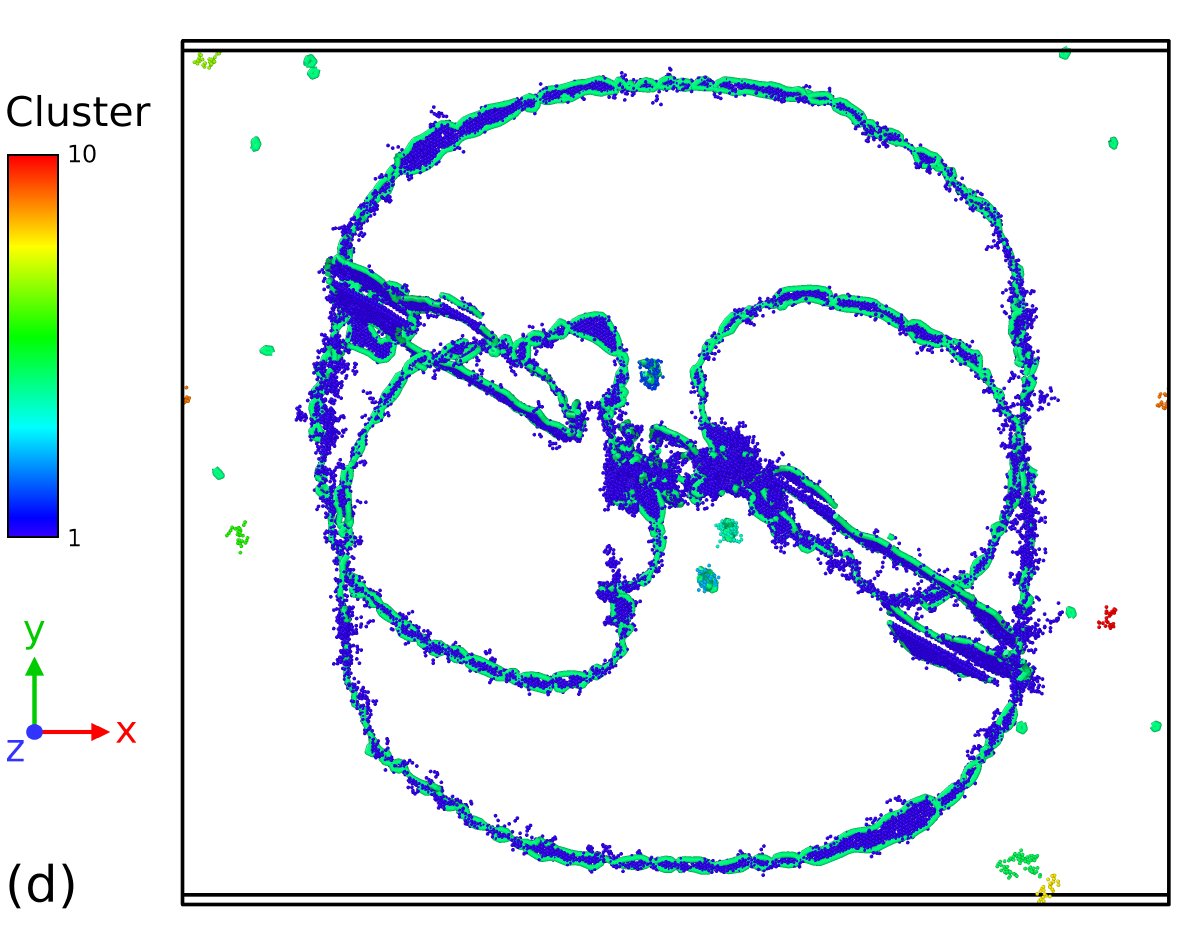}
  \caption{MD simulation snapshots after 14 ps of simulation and structural analysis using PTM \textbf{(a)} or only classifier SL-CSA based on logistic regression \textbf{(b)}. Additionally, an extra step was performed to remove noisy atoms by applying the Cluster Analysis in \texttt{OVITO} to the PTM results \textbf{(c)} and using the Mahalanobis distance criterion from the SL-CSA procedure \textbf{(d)}. Dislocation lines extracted using DXA are superimposed in \textbf{(d)}.
  }
  \label{fig:snapshots_FR_loop}
\end{figure*}
\subsection{Frank-Read source in Aluminium at 700 K}
In this example, we investigate the capabilities of the classifier to extract the atoms that belong to defect structures from the bulk, and, in particular, to identify atoms belonging to dislocation cores. We emphasize that the classifier was not trained to distinguish any defective configuration.
Hence, the analysis developed below only concerns the identification of crystalline and non-crystalline (or defective) atoms.

The present simulation involves the double emission of Frank-Read dislocation sources in FCC Al at high temperature, an example that has been previously used  (at low temperature) to demonstrate the capability of the PTM algorithm in~\cite{Stukowski_2010}, with a setup similar to the one described in~\cite{Koning_2003}.
The only difference is that the dislocation sources have been pinned by two cylindrical pores periodically along the $z$ direction. The system with 2,300,504 atoms was initially equilibrated in the NPT ensemble at 0 GPa and 700 K for 10 ps, using similar coupling constants as for the HCP to BCC simulation described in the previous section.
Finally, a shear stress $\sigma_{yz}$ of 1.5 GPa was applied to the simulation cell,  while keeping the temperature at 700 K in order to trigger the expansion of dislocation lines emerging from the Frank-Read sources. A snapshot of the simulation cell at t = 14 ps is displayed in Figure~\ref{fig:snapshots_FR_loop}.

The aim here is to investigate the capability of the SL-CSA procedure to extract defects, i.e. atoms that have not been assigned by the classifier to a specific crystal structure. Figure~\ref{fig:snapshots_FR_loop}-\textbf{a}, \textbf{b} depicts the microstructure analyzed using PTM0.1 and SL-CSA, after removing atoms belonging to the FCC structure. 
The significant thermal noise in the simulation cell causes the presence of atomic environments deviating from the ideal FCC structure, including non-crystalline and BCC atoms.
Here, PTM tends to identify more non-crystalline atoms than SL-CSA, the latter leading to the identification of HCP atoms in the FCC bulk. Since FCC Al tends to easily nucleate HCP stacking faults, the fact that thermally disturbed FCC bulk atoms are being identified as HCP is not that uncommon. A main difference subsists between both classifiers: PTM method appears more dependent on the local thermal/stress state than SL-CSA. 
Indeed, the area that is relaxed by the dislocation loop (i.e. at lower stress) appears less noisy than the one that is not in its vicinity, in opposition to the SL-CSA, where the noisy atoms look homogeneously distributed across the sample.
In the end, what really matters is the ability of the present methods to extract defective parts of atomistic simulations for subsequent analysis. For example, defective atoms identified with SL-CSA or PTM could be fed to the Dislocation eXtraction Analysis (DXA) tool~\cite{Stukowski2012dxa} for dislocation identification.
Both PTM and SL-CSA seem able to identify the dislocation loop stacking faults generated by the Frank-Read sources, constituted by atoms assigned with the HCP crystal structure. Employing the cluster analysis available in \texttt{OVITO} allows for removing noisy bulk atoms and the remaining defective structure extracted from both PTM and SL-CSA classifications are displayed in Figure~\ref{fig:snapshots_FR_loop}-\textbf{c}, \textbf{d} respectively. In comparison to the PTM0.1 analysis workflow, SL-CSA looks more robust and leads to the extraction of almost only dislocation core atoms. Such a structure would be a good candidate for performing extended analysis in terms of dislocation density calculations. However, dislocation line extraction is not the object of the present work and will be part of further studies. Overall, our crystal structure classification procedure SL-CSA performs rather well compared to the existing tools from the literature, even when both non-hydrostatic stresses and high temperature are involved such as in this dislocation-mediated plasticity simulation toy model.

\section{Conclusion and perspectives}
We have introduced a novel classifier that surpasses the capabilities of conventional approaches, such as a-CNA and PTM, when it comes to identifying crystal structures under extreme conditions like high temperature, high pressure, and large deformation. This makes our method particularly suitable for real-time analysis of molecular dynamics (MD) simulations.

Our proposed classifier operates on a simple learning process, utilizing a training database that encompasses various structures of interest, including BCC, FCC, HCP, and c-DIA. The characterization of local structures is facilitated by a spectral descriptor, which captures the geometric arrangement of neighboring atoms and represents it as a vector.
We have proposed a modification to the conventional bispectrum descriptor, ensuring that a fixed number of neighbors is incorporated within the descriptor. This modification enables the analysis of materials at varying densities without loosing accuracy. The newfound insensitivity of the modified descriptor to density changes has a significant impact on the size of the required training database for the classifiers, while maintaining transferability.
The training database includes configurations at high temperatures and small deformations, i.e. in the elastic regime. 

A simple logistic regression is employed for the classifier, carefully controlling the balance between false positives and false negatives. The classifier is applied following dimensionality reduction using an LDA discriminator. While LDA may lose information about the covariance matrix differences between the classes, we mitigate this by refining the results with a test based on Mahalanobis distance within each and across the classes.
We compare the performance of our current SL-CSA classifier to standard classification tools (a-CNA and PTM) across various densities, temperatures, and deformations. Notably, the SL-CSA classifier demonstrates superior reliability, even in scenarios where temperature and deformation interact.

Finally, the SL-CSA classifier was examined on large-scale simulations of solid-solid phase transformations and the detection of dislocation core atoms. These simulations showed that our classifier can conduct an analysis of crystalline structures with higher precision than traditional techniques, allowing to accurately estimate a proportion of atoms that belong to a given crystalline structure. We are optimistic that this capability will be advantageous for raising the accuracy of coarse-grained models of such processes. 

In perspective, given its transferability and capability to analyze unknown features, the present method holds potential for further expansion in identifying more complex crystalline structures and directly detecting specific defect types, such as dislocation cores. Additionally, its application can extend to novelty detection in the field of materials science, such as recent nano-phases inclusions \cite{goryaeva_compact_2023}.

\section*{Data Availability}
Data will be made available on request.


\end{document}